\documentclass[journal,twoside,web]{ieeecolor}
\usepackage{tmi}
\usepackage{cite}
\usepackage{amsmath,amssymb,amsfonts}
\usepackage{algorithm}
\usepackage{graphicx}
\usepackage{textcomp}
\usepackage{booktabs}       
\usepackage{multirow}

\usepackage{caption}
\usepackage{subcaption}

\usepackage{algorithm, algorithmicx, algpseudocode}

\def\BibTeX{{\rm B\kern-.05em{\sc i\kern-.025em b}\kern-.08em
		T\kern-.1667em\lower.7ex\hbox{E}\kern-.125emX}}
\renewcommand{\figurename}{FIGURE} 
\renewcommand{\thefigure}{\arabic{figure}} 
\begin{document}

\title{Domain-conditioned and Temporal-guided Diffusion Modeling for Accelerated Dynamic MRI Reconstruction}
\author{Liping Zhang, Iris Yuwen Zhou, Sydney B. Montesi, Li Feng, and Fang Liu
\thanks{This work was supported by the National Institute of Biomedical Imaging and Bioengineering (R21EB031185), the National Institute of Arthritis and Musculoskeletal and Skin Diseases (R01AR081344, R01AR079442, and R56AR081017), and the National Heart Lung and Blood Institute (K23HL150331 and K25HL148837). (Corresponding author: Fang Liu)}
\thanks{Liping Zhang, Iris Yuwen Zhou, and Fang Liu are with the Athinoula A. Martinos Center for Biomedical Imaging, Massachusetts General Hospital and Harvard Medical School, Charlestown, MA, USA. (e-mail: lpzhang90@mgh.harvard.edu; iris.zhou@mgh.harvard.edu; fliu12@mgh.harvard.edu)}
\thanks{Sydney B. Montesi is with the Division of Pulmonary and Critical Care Medicine, Massachusetts General Hospital and Harvard Medical School, Boston, MA, USA. (e-mail: sbmontesi@mgb.org)}
\thanks{Li Feng is with the Center for Advanced Imaging Innovation and Research, Department of Radiology, New York University Grossman School of Medicine, New York City, New York, USA. (li.feng@nyulangone.org)
}}

\maketitle

\begin{abstract}
	\textbf{Purpose:} To propose a domain-conditioned and temporal-guided diffusion modeling method, termed dynamic Diffusion Modeling (dDiMo), for accelerated dynamic MRI reconstruction, enabling diffusion process to characterize spatiotemporal information for time-resolved multi-coil Cartesian and non-Cartesian data.
    \textbf{Methods:} The dDiMo framework integrates temporal information from time-resolved dimensions, allowing for the concurrent capture of intra-frame spatial features and inter-frame temporal dynamics in diffusion modeling. It employs additional spatiotemporal ($x$-$t$) and self-consistent frequency-temporal ($k$-$t$) priors to guide the diffusion process. This approach ensures precise temporal alignment and enhances the recovery of fine image details. To facilitate a smooth diffusion process, the nonlinear conjugate gradient algorithm is utilized during the reverse diffusion steps. The proposed model was tested on two types of MRI data: Cartesian-acquired multi-coil cardiac MRI and Golden-Angle-Radial-acquired multi-coil free-breathing lung MRI, across various undersampling rates.
    \textbf{Results:} dDiMo achieved high-quality reconstructions at various acceleration factors, demonstrating improved temporal alignment and structural recovery compared to other competitive reconstruction methods, both qualitatively and quantitatively. This proposed diffusion framework exhibited robust performance in handling both Cartesian and non-Cartesian acquisitions, effectively reconstructing dynamic datasets in cardiac and lung MRI under different imaging conditions.
    \textbf{Conclusion:} This study introduces a novel diffusion modeling method for dynamic MRI reconstruction.
\end{abstract}



\section{Introduction}
\IEEEPARstart{M}{agnetic} resonance imaging (MRI) is a non-invasive imaging modality extensively used in medical diagnostics and clinical research due to its ability to provide detailed anatomical and functional information. However, limited by its acquisition nature, MRI is considered a slow imaging modality compared to others. Accelerated MRI has been developed to enable undersampling k-space to reduce scan time while maintaining favorable image quality by combining advanced image reconstruction algorithms. Classic methods such as parallel imaging \cite{sodickson1997simultaneous, pruessmann1999sense, griswold2002grappa}, compressed sensing \cite{lustig2007sparse}, and low-rank methods \cite{lingala2011accelerated, christodoulou2013high, haldar2013low, dong2014compressive, shin2014calibrationless, otazo2015low, zhang2015accelerating, he2016accelerated} have demonstrated promises in maintaining desirable image reconstruction quality using undersampled data focusing on specific algorithm design. 

Deep learning has become another disruptive method for improving MRI reconstruction, leveraging a data-driven approach that learns necessary image features from the data itself to recover undersampled k-space information. Various deep learning architectures, such as convolutional neural network (CNN), recurrent neural network, generative adversarial network, and transformer-based model, have been employed to reconstruct undersampled k-space data with successful performance \cite{wang2016accelerating, sun2016deep, yang2017dagan, aggarwal2018modl, hammernik2018learning, eo2018kiki, zhu2018image, liu2019santis, ahmad2020plug, akccakaya2019scan, sriram2020end}. In addition, deep learning can also be used to reconstruct image data with multi-dimensional information, such as in dynamic MRI and multi-parametric quantitative MRI (qMRI), with specific algorithm design to characterize motion and contrast changes during dynamic imaging \cite{schlemper2017deep, qin2018convolutional, qin2021complementary, kustner2020cinenet, dar2020prior, yoo2021time, kleineisel2022real, zhang2024camp, phair2024motion, catalan2025unsupervised} and signal variations during qMRI modeling \cite{liu2019mantis, liu2020high, liu2021magnetic, jun2021deep, bian2024improving, jun2024zero}.

Recently, a new emerging class of deep learning, generative AI, has gained significant attention. Diffusion modeling as a generative AI approach has been found effective, efficient, and robust in various imaging applications for image denoising, deartifacting, and synthesis for natural images \cite{Ho2020ddpm, nichol2021improved, dhariwal2021diffusion}. Recent studies have extended diffusion modeling for MRI reconstruction, highlighting its potential to enable a faithful high-quality image restoration for undersampled k-space data \cite{chung2022score, chung2022come, gungor2023adaptive, luo2023bayesian, bian2024diffusion}. In diffusion modeling, the forward diffusion process gradually introduces noise into the image data following a predefined scheme, while the reverse diffusion process recovers the desired data from random noise through learned Gaussian transitions. While unconditioned diffusion modeling is well suited for natural image generation, as its name noted as generative AI, the diffusion process conditioned on the acquired k-space data can be used to enforce data consistency and incorporate MR physics constraints into diffusion modeling-based MRI reconstruction. For example, the \underline{Di}ffusion \underline{Mo}deling with domain-conditioned prior guidance (DiMo) method has been shown to improve performance not only on the static MR images but also on the parametric maps in quantitative MRI reconstruction \cite{bian2024diffusion}. This method introduced the domain-conditioned diffusion model within the frequency and parameter domains where the prior MRI physics are used as embeddings in the diffusion model, enforcing data consistency to guide the diffusion process, characterizing k-space encoding in MRI reconstruction, and leveraging MR signal modeling for qMRI reconstruction.

In this study, we introduce a new approach called \underline{d}ynamic \underline{Di}ffusion \underline{Mo}deling (dDiMo), which features innovative domain-conditioned and temporal-guided elements for reconstructing accelerated dynamic MRI data. The dDiMo framework builds upon the original DiMo approach \cite{bian2024diffusion} by integrating temporal information from additional time-resolved dimensions. This enhancement allows it to capture rapid temporal variations in multi-coil dynamic MRI data. Unlike other diffusion methods \cite{chung2022score, gungor2023adaptive, luo2023bayesian, bian2024diffusion}, which focus on spatial information within individual frames during the denoising process, dDiMo simultaneously considers both intra-frame spatial features and inter-frame temporal dynamics. This dual capability enables dDiMo to learn temporally-aware denoising priors by utilizing the temporal coherence of adjacent time frames, facilitating joint noise removal. Additionally, dDiMo harnesses spatiotemporal ($x$-$t$) and self-consistent frequency-temporal ($k$-$t$) priors derived from time-resolved data to guide the diffusion process. The framework also incorporates the nonlinear conjugate gradient (CG) algorithm \cite{lustig2007sparse} into the reverse diffusion steps to further improve the performance of this diffusion process. The versatility and effectiveness of dDiMo are demonstrated through experiments conducted on both Cartesian-acquired multi-coil cardiac MRI and continuously acquired free-breathing Golden-Angle-Radial multi-coil lung MRI. The results indicate that dDiMo consistently produces high-quality reconstructions across various undersampling rates and imaging conditions, underscoring its robustness and adaptability for dynamic MRI.

The rest of the paper is organized as follows: Section 2 provides the methods for dDiMo, Section 3 describes the experiment settings, Section 4 presents the experiment results, Section 5 discusses the method's limitations, and Section 6 concludes the paper. The appendix provides additional results for ablation studies in this work.

\section{Methods}
\subsection{Dynamic MRI Reconstruction}
Accelerated dynamic MRI involves the following measurement model:
\begin{equation}
    \tilde{y} = \mathcal{A}x + \varepsilon,
    \label{eq:forward_process}
\end{equation}
where
$x \in \mathbb{C}^N$ represents the MR image sequence of interest, consisting of $N$ pixels in the spatiotemporal domain ($x$-$y$-$t$ space, also denoted as $x$-$t$), $\tilde{y} \in \mathbb{C}^{N_cM}$ is the corresponding undersampled k-space measurement, acquired with $N_c$ coils, where each coil contains $M$ sampled data points in the frequency-temporal domain ($k_x$-$k_y$-$t$ space, also denoted as $k$-$t$), and $\varepsilon \in \mathbb{C}^{N_cM}$ is the measurement noise. The forward measurement operator $\mathcal{A}: \mathbb{C}^N \to \mathbb{C}^{N_cM}$ is defined as:
\begin{equation}
    A := \mathcal{MFS},
    \label{eq:forward_operator}
\end{equation}
where $\mathcal{S}: \mathbb{R}^N \to \mathbb{C}^{N_cN}$ is the coil sensitivity maps of $N_c$ coils, $\mathcal{F}: \mathbb{C}^{N_cN} \to \mathbb{C}^{N_cN}$ is the Fourier transform, and $\mathcal{M}: \mathbb{C}^{N_cN} \to \mathbb{C}^{N_cM} (M \ll N)$ is the binary undersampling mask corresponding to $M$ sampled data points. The reconstruction is an ill-posed inverse problem and can be formulated as solving the following optimization problem:
\begin{equation}
\hat{y} = \arg\mathop{\min}\limits_y \|\mathcal{M}y-\tilde{y} \|^2_2 + \lambda_{xt}\mathcal{R}_\text{xt}(\mathcal{S}^{H}\mathcal{F}^{H}y) + \lambda_{kt}\mathcal{R}_\text{kt}(y),
\end{equation}
where $\mathcal{R}_\text{xt}$ represents a regularization term in the $x$-$t$ domain and $\mathcal{R}_\text{kt}$ is a regularization term in the $k$-$t$ domain. The parameters $\lambda_\text{xt}$ and $\lambda_\text{kt}$ control the trade-off between the data fidelity term and the regularization terms in those domains.

The nonlinear CG method can be used to optimize $\hat{y}$ iteratively \cite{lustig2007sparse}. Specifically, at iteration $k$, we can apply the following update:
\begin{equation}
    \hat{y}_{k+1} = \hat{y}_{k} + \alpha_k d_{k},
    \label{eq:cg_step}
\end{equation}
where the positive step size $\alpha_k$ is obtained by a line search, and the search direction $d_{k}$ is
generated by:
\begin{equation}
    d_{k+1} = -g_{k+1} + \beta_k d_{k},
    \label{eq:cg_direction}
\end{equation}
where
$g_{k}$ is the gradient for $\hat{y}_{k}$ at iteration $k$, and $\beta_k$ is the CG update parameter defined as:
\begin{equation}
    \beta_k = \frac{g^T_{k+1} g_{k+1}}{g^T_{k} g_{k}}.
\end{equation}

\subsection{dDiMo for Dynamic MRI Reconstruction}

\begin{figure*}
	\centering
	\includegraphics[width=\textwidth]{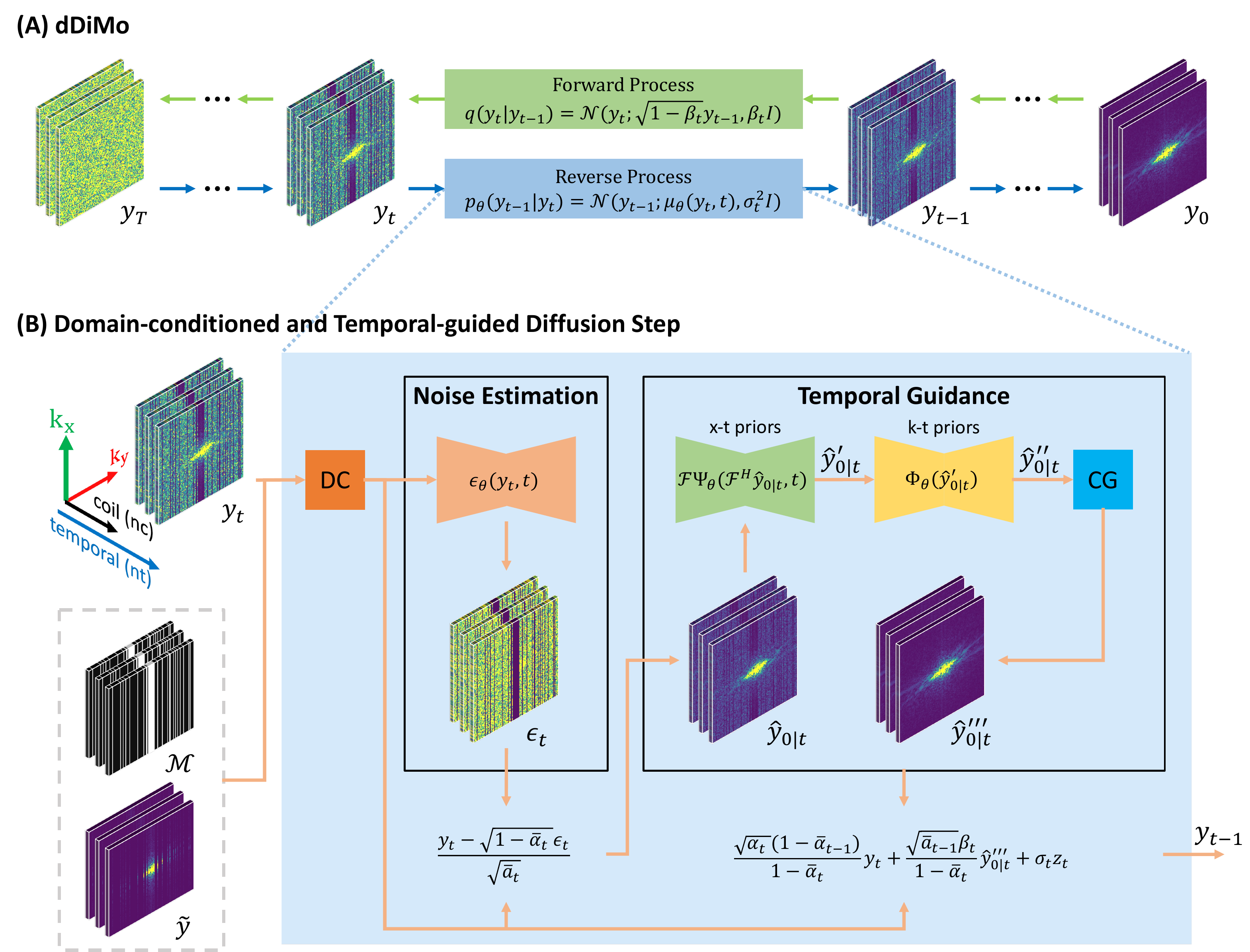}
	\caption{
    The overall framework of the proposed dDiMo, which integrates temporal information into the diffusion process for dynamic MRI reconstruction.
(A) dDiMo performs progressive diffusion (forward process) or sampling (reverse process) of multi-coil dynamic k-space frames in the k-space domain, utilizing domain-conditioned and temporal-guided diffusion steps.
(B) Each diffusion step incorporates a data consistency layer and includes several key components to adapt to dynamic image reconstruction:
(1) A 3D CNN-based noise estimation network $\epsilon_{\theta}$ that captures spatial features within frames and temporal dynamics across frames, leveraging temporal coherence to learn denoising priors sensitive to temporal changes.
(2) A 3D CNN-based $x$-$t$ network $\Psi_{\theta}$ that models temporal dynamics in the spatiotemporal image domain from intermediate denoised results $\hat{y}_{0|t}$, improving temporal alignment.
(3) A 3D CNN-based $k$-$t$ network $\Phi_{\theta}$ that enforces MRI physics constraints through self-consistency learning in the ACS region of $k$-$t$ space, refining the denoising process and ensuring k-space consistency.
(4) A nonlinear CG module that iteratively refines the reverse diffusion steps, adhering to physical constraints and enhancing robustness under challenging conditions.
    }
    \label{fig:dDiMo}
\end{figure*}

Inherited from the original DiMo method, the dDiMo framework features a forward diffusion process that introduces noise into k-space, followed by a reverse diffusion process that removes the noise to recover data. Unlike the original DiMo, which reconstructs static images, dDiMo incorporates additional temporal information from time-resolved dimensions, allowing it to capture dynamic features and enhance image reconstruction for dynamic MRI data. An illustrative diagram of the proposed dDiMo framework is presented in Figure~\ref{fig:dDiMo}. This new framework includes several key components designed to adapt to the reconstruction of dynamic image data: (1) A 3D CNN-based noise estimation network that characterizes spatial features within frames and temporal dynamics across frames, leveraging temporal coherence to learn denoising priors aware of temporal changes. (2) A 3D CNN-based $x$-$t$ network that captures temporal dynamics in the spatiotemporal image domain from intermediate denoised results, guiding the diffusion process toward better temporal alignment. (3) A 3D CNN-based $k$-$t$ network that enforces MRI physics constraints through data self-consistency learning within the ACS region of $k$-$t$ space, refining the denoising process and ensuring consistency within k-space. (4) A nonlinear CG module that iteratively refines the reverse diffusion steps, maintaining adherence to physical constraints and enhancing robustness. Further details about each component will be provided in the following sections.

\subsubsection{Diffusion modeling for dynamic data}
\textit{Forward Diffusion}: The forward diffusion process in dDiMo is modeled as a Markov chain that progressively adds Gaussian noise to a dynamic sequence of fully sampled multi-coil k-space data, denoted as $y_0$. This dynamic sequence includes multiple adjacent time frames, and the process is conditioned on the acquired measurements $\tilde{y}$, which also span multiple time frames, along with the corresponding undersampling mask $\mathcal{M}$.
The forward diffusion process transforms the initial distribution $q(y_0)$ into a sequence of increasingly noisy samples, which can be mathematically represented as:
\begin{equation}
    q(y_{1:T}| y_0, \tilde{y}, \mathcal{M})  := \prod^T_{t=1} q(y_{t}| y_{t-1}, \tilde{y}, \mathcal{M}),
\end{equation}
\begin{equation}
    q(y_{t}| y_{t-1}, \tilde{y}, \mathcal{M}) := \mathcal{N}(y_{t}; \sqrt{ 1- \beta_t} y_{t-1}, \beta_t I ),
\end{equation}
where $t$ is the diffusion step index, $T$ is the total number of diffusion steps, $I$ is an identity matrix, and the noise scaling sequence $\beta_t \in [0,1)$ defines the magnitude of noise variance introduced at $t$ step. By leveraging the properties of the Gaussian distribution, direct sampling can be formulated at any timestep $t$ given the fully-sampled data $y_0$ as follows:
\begin{equation}
    q(y_{t}| y_{0}, \tilde{y}, \mathcal{M}) := \mathcal{N}(y_{t}; \sqrt{\bar{\alpha}_t} y_{0}, (1-\bar{\alpha}_t) I),
\end{equation}
where $\alpha_t = 1 - \beta_t$ and $\bar{\alpha}_t = \prod ^t _{i=1}\alpha_i$. As the forward process progresses and $\bar{\alpha}_T \rightarrow 0$, the distribution
$q(y_{T}| y_{0}, \tilde{y}, \mathcal{M})$ approaches an isotropic Gaussian distribution:
\begin{equation}
    q(y_{T}| y_{0}, \tilde{y}, \mathcal{M}) \approx \mathcal{N}(y_{T}; 0, I).
\end{equation}

Unlike static image data, this diffusion modeling processes several adjacent time frames at the same time, allowing it to utilize temporal information across frames during the training of the denoising network. This approach helps maintain better temporal alignment throughout the reconstruction. By taking inter-frame relationships into account, the dynamics of the forward diffusion process are influenced by both temporal and spatial factors. This enables the model to capture the inherent spatial structure and the temporal evolution within the dynamic series effectively.

\textit{Reverse Diffusion}: In the reverse diffusion process, the multi-coil dynamic k-space series $y_0$ is generated by initializing from random Gaussian noise $y_T \sim \mathcal{N}(0,I)$ and progressively removing the estimated noise over $T$ steps. This process can be formalized as:
\begin{equation}
    p_\theta(y_{0:T}|\tilde{y}, \mathcal{M})  := p_\theta(y_T |\tilde{y}, \mathcal{M}) \prod_{t=1}^{T} p_\theta(y_{t-1}| y_{t}, \tilde{y},\mathcal{M}),
\end{equation}
\begin{equation}
    p_\theta(y_{t-1}| y_{t}, \tilde{y}, \mathcal{M}) := \mathcal{N} (y_{t-1}; \mu_\theta(y_t, t, \tilde{y}, \mathcal{M}), \sigma_t^2 I),
\end{equation}
where $\sigma_t^2 = \frac{1-\bar{\alpha}_{t-1}}{1-\bar{\alpha}_t} \beta_t$. In practice, a CNN can be used to predict the noise directly rather than to predict the mean \cite{Ho2020ddpm}. This can be expressed as:
\begin{equation}
    \mu_\theta (y_t, t, \tilde{y}, M) = \frac{1}{\sqrt{\alpha_t}} (y_t - \frac{\beta_t}{1 - \bar{\alpha}_t}\epsilon_\theta(y_t, t)),
\end{equation}
where $\epsilon_\theta$ denotes the noise prediction model using a CNN, for example, a U-Net \cite{nichol2021improved, dhariwal2021diffusion}. To capture the spatiotemporal features of the dynamic data, we applied 3D convolutional layers (i.e., 2D spatial + 1D temporal) in a 3D U-Net (Supporting Information Figure 6) to extract both intra-frame and inter-frame information, enabling the model to exploit the underlying spatial-temporal coherence of the data for more effective noise removal. Additionally, like DiMo \cite{bian2024diffusion}, data consistency (DC) is incorporated into estimating $y_t$, where a linear combination of partially scanned data $\tilde{y}$ is added to $y_t$ to ensure data fidelity. 

To construct a desired k-space data from the noise, the reverse process can be written as:
\begin{equation}
y_{t-1} = \frac{1}{\sqrt{\alpha_t}} (y_t - \frac{\beta_t}{1 - \bar{\alpha}_t}\epsilon_\theta(y_t, t)) + \sigma_{t}z_t,
\label{eq:reverse_process}
\end{equation}
where $\epsilon_{\theta}\left(y_t ,t\right)$ represents the predicted noise at step $t$, $z_t \sim \mathcal{N}(\mathbf{0},\mathbf{I})$ denotes a random variable sampled from the normal distribution. The $y_0$ can be obtained by repeating the reverse step $T$ times recursively. However, this standard diffusion process has not incorporated explicit modeling in the $x$-$t$ and $k$-$t$ domains, which can be further utilized to characterize temporal features, as described in the following sessions.

\subsubsection{$x$-$t$ priors for spatiotemporal learning}
Like conventional reconstruction to leverage regularization in the $x$-$t$ domain, we further investigated the characterization of temporal image features by explicitly learning from the data in the $x$-$t$ domain. This is achieved using a 3D CNN to learn the associated $x$-$t$ priors based on the intermediate denoised image sequences. Although it seems intuitive to directly learn these features from the intermediate results $y_t$ in Eq.~\eqref{eq:reverse_process} during the reverse process, applying $x$-$t$ prior learning directly to $y_t$ can lead to suboptimal performance. This is primarily due to the substantial noise present in the early stages of the reverse process, which can obscure the true temporal features. To address this challenge, we propose a strategy for learning $x$-$t$ priors based on ``clean" data. Inspired by previous work in diffusion \cite{lin2025diffbir}, we can isolate the underlying ``clean" data $\hat{y}_{0|t}$ from the noisy data $y_t$ using the following approach:

\begin{equation}
    \hat{y}_{0|t} = \frac{1}{\sqrt{\bar{\alpha}_t}}\left(y_t - \sqrt{1-\bar{\alpha}_t}\epsilon_{\theta}\left(y_t ,t\right)\right),
    \label{eq:reverse_process_part1}
\end{equation}

This enables the direct operation of the ``clean" data without worrying about associated noises. Specifically, this $k$-$t$ prior learning process can be formulated as:
\begin{equation}
    \hat{y}_{0|t}^{'} = \mathcal{F}\Psi_{\theta}(\mathcal{F}^{H}\hat{y}_{0|t}, t),
    \label{eq:xt}
\end{equation}
where $\mathcal{F}$ represents a Fourier transform applied to the estimated clean k-space data $\hat{y}_{0|t}$ to convert it into $x$-$t$ space, and $\mathcal{F}^H$ denotes the inverse Fourier transform operation. The $\Psi_{\theta}$ represents a 3D CNN that captures the temporal dynamics.
, where, in this study, we employ a 3D U-Net (Supporting Information Figure 7) for $\Psi_{\theta}$.

\subsubsection{$k$-$t$ priors for data self-consistency}
\begin{figure}
	\centering
	\includegraphics[width=0.5\textwidth]{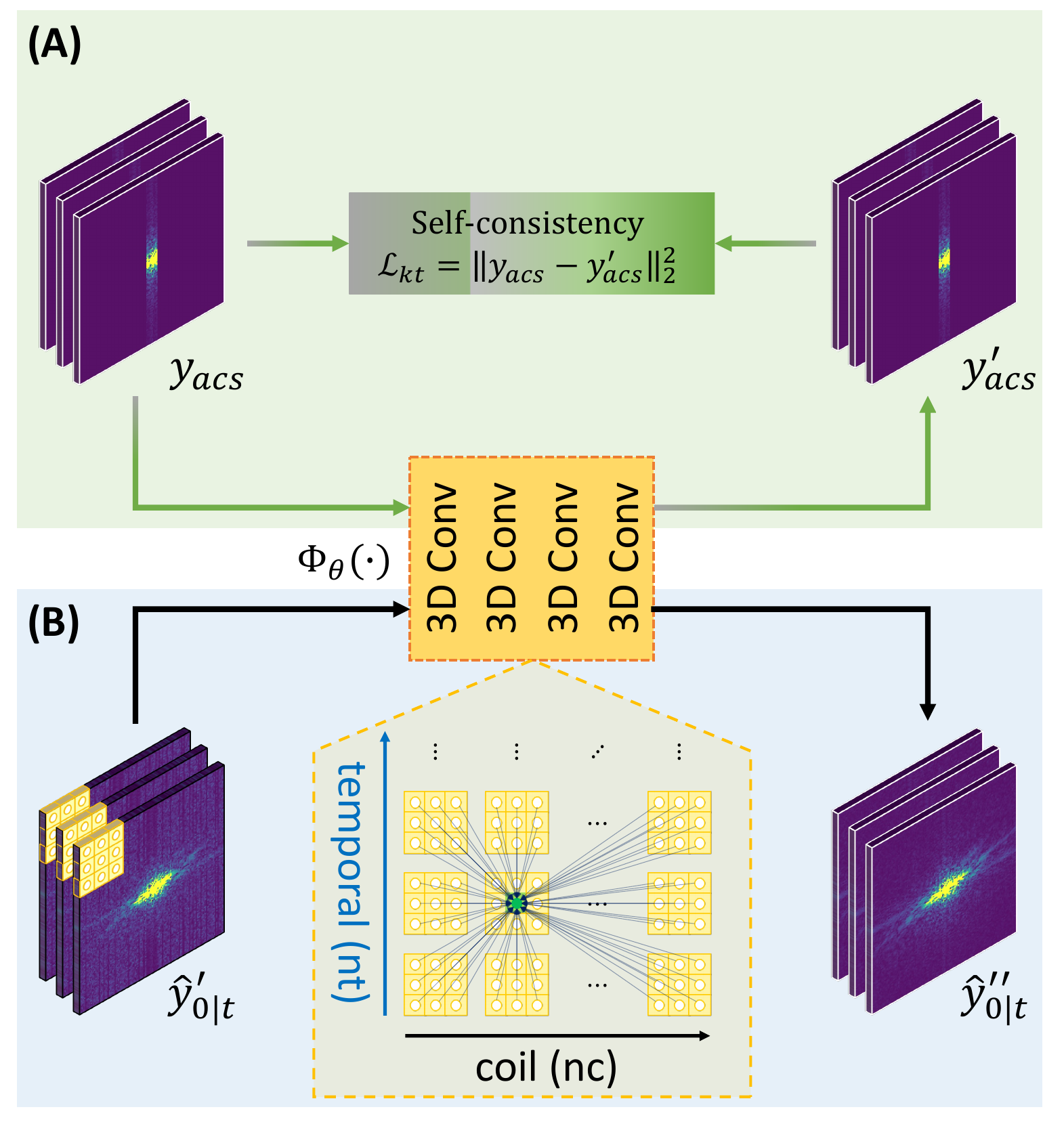}
	\caption{
    Schematic illustration of the use of $k$-$t$ priors during inference and training. (A) $k$-$t$ priors are derived from the ACS data in the $k$-$t$ space during training, guided by a self-consistency loss. (b) Self-consistency $k$-$t$ priors are applied to the input $\hat{y}_{0|t}^{'}$ during inference to enforce consistency in the frequency-temporal domain.
    }
    \label{fig:dDiMo-kt}
\end{figure}

Shift-invariant k-space correlations from fully-sampled auto-calibration signal (ACS) regions have been widely used for k-space interpolation and as regularization priors in MRI reconstruction \cite{griswold2002grappa, akccakaya2019scan, zhang2024camp}. In dynamic imaging, similar $k$-$t$ patterns can be extracted from ACS data across consecutive time frames \cite{huang2005k, zhang2023k}. Inspired by the work from scan-specific neural networks for k-space interpolation\cite{akccakaya2019scan} and our previous k-space implementation \cite{zhang2023k, zhang2024camp}, we utilize a 3D CNN with four convolutional layers, each with a kernel size of 3$\times$3$\times$3, as shown in Figure~\ref{fig:dDiMo-kt}, to capture $k$-$t$ correlations through a self-consistency learning mechanism during the training phase. This process involves feeding ACS data from multiple time frames into the network, enabling it to learn the underlying $k$-$t$ patterns by predicting the input data itself as accurately as possible, expressed mathematically as:
\begin{equation}
y_{acs}^{'} = \Phi_{\theta}\left(y_{acs}\right),
\end{equation}
where $\Phi_{\theta}$ denotes the 3D CNN that encodes the $k$-$t$ priors from the ACS data,  $y_{acs}$ represents the ACS data used to capture the self-consistency $k$-$t$ priors, and $y_{acs}^{'}$ denotes the corresponding network prediction of $y_{acs}$.  Once the training is complete, the learned priors can then be incorporated into the reverse diffusion to enforce the temporal frequency consistency of the k-space data recovery at each diffusion step. This process can be described as:
\begin{equation}
\hat{y}_{0|t}^{''} \leftarrow \Phi_{\theta}\left(\hat{y}_{0|t}^{'}\right),
\label{eq:kt}
\end{equation}
where $\hat{y}_{0|t}^{'}$ is the result of Eq.~\eqref{eq:xt}. A schematic illustration of the use of $k$-$t$ priors during training and inference is shown in Figure~\ref{fig:dDiMo-kt}. 

\subsubsection{Nonlinear conjugate gradient layer}
Followed by the $x$-$t$ and $k$-$t$ feature characterization through CNNs, a layer implementing the iterative nonlinear CG is applied to refine the result in each of the reverse diffusion steps. This iterative refinement using CG enhances the robustness of the reverse diffusion process to noise perturbations. It also complements the automatic learning process by reinforcing well-known image features, such as temporal sparsity, into the diffusion process. The CG layer is formulated as:
\begin{equation}
    \hat{y}_{0|t}^{'''} = \mathbf{CG} \left( \hat{y}_{0|t}^{''} \right),
    \label{eq:cg}
\end{equation}
where $\hat{y}_{0|t}^{''}$ is obtained from Eq. \eqref{eq:kt}. Each step of the CG optimization process consists of two primary updates, as defined in Eq.\eqref{eq:cg_step} and Eq.\eqref{eq:cg_direction}, aimed at minimizing the following objective function:
\begin{equation}
\mathcal{L}(\hat{y}_{0|t}^{'''})
= \| \mathcal{M} \hat{y}_{0|t}^{'''} - \tilde{y} \|_2^2 + \lambda_{td} \|\nabla_t (\mathcal{S}^H \mathcal{F}^H \hat{y}_{0|t}^{'''})\|_1,
\label{eq:cg_objective}
\end{equation}
where $\nabla_t(\cdot)$ represents temporal finite differences applied along dynamic dimensions to enforce temporal sparsity in the reconstructed dynamic images and $\lambda_{td}$ is a weighting factor \cite{feng2014golden, feng2016xd}.

Once the temporal-guided ``clean" data $\hat{y}_{0|t}^{'''}$ has been obtained through $x$-$t$ and $k$-$t$ regularization and non-linear CG refinement, it can be converted back into its original state at the diffusion step $t-1$ as:
\begin{equation}
    y_{t-1} 
    =\frac{\sqrt{\bar{\alpha}_t}\left(1-\bar{\alpha}_{t-1}\right)}{1-\bar{\alpha}_t}y_t
    + \frac{\sqrt{\bar{\alpha}_{t-1}}\beta_t}{1-\bar{\alpha}_t}\hat{y}_{0|t}^{'''}
    + \sigma_{t}z_{t}
\label{eq:bk}
\end{equation}

The final $y_0$ can be obtained by repeating the reverse diffusion step $T$ times recursively following Eq.\eqref{eq:bk}.

\subsubsection{Overall training strategy}

During training, the noise estimation network $\epsilon_{\theta}$, $x$-$t$ network $\Psi_{\theta}$, and $k$-$t$ network $\Phi_{\theta}$ are jointly optimized by minimizing the following objective function:
\begin{equation}
    \mathcal{L} 
    = \| \epsilon  - \epsilon_{\theta}\left(y_t,t\right) \|^2_2
    + \lambda_{xt} \| y_0  - \hat{y}_{0|t}^{'}\|^2_2
    + \lambda_{kt} \| y_{acs} - y_{acs}^{'} \|^2_2
    \label{eq:loss_function}
\end{equation}
where the terms $\lambda_{xt}$, and $\lambda_{kt}$ are weighting factors that balance the trade-offs between noise estimation, $x$-$t$ prior and $k$-$t$ prior learning. The pseudo-algorithms for the training and reverse sampling processes of dDiMo are provided in Alg.\ref{alg:train} and Alg.\ref{alg:sample}, respectively.

\begin{algorithm}
\caption{Training Process of dDiMo}\label{alg:train}
\begin{algorithmic}[1]
\Repeat
\State $y_0 \sim q\left(y_0\right)$, $\mathcal{M} \sim \Omega\left(\mathcal{M}\right)$
\State $\tilde{y} = \mathcal{M}y_0$, $y_{acs} = \mathcal{P}\left(\tilde{y}\right)$
\State $t \sim \text{Uniform}\left(\{1,\cdots,T\}\right)$
\State $\epsilon \sim \mathcal{N}\left(\mathbf{0},\mathbf{I}\right)$
\State $y_t \leftarrow \sqrt{\bar{\alpha}_t}y_0 + \sqrt{1-\bar{\alpha}_t}\epsilon$
\State $y_t \leftarrow \mathcal{M}\left(\lambda_t\tilde{y} + \left(1-\lambda_t\right)y_t\right) + \left(1-\mathcal{M}\right)y_t$
\hfill$\triangleright$ DC
\State $\hat{y}_{0|t} = \frac{1}{\sqrt{\bar{\alpha}_t}}\left(y_t - \sqrt{1-\bar{\alpha}_t}\epsilon_{\theta}\left(y_t,t\right)\right)$
\State $\hat{y}_{0|t}^{'} = \mathcal{F}\Psi_{\theta}\left(\mathcal{F}^{H}\hat{y}_{0|t},t\right)$
\hfill$\triangleright$ $x$-$t$ prior
\State $y_{acs}^{'} = \Phi_{\theta}\left(y_{acs}\right)$
\hfill$\triangleright$  $k$-$t$ prior
\State Take gradient descent update step
\State $\quad\quad
\begin{aligned}[t]
&\nabla_{\epsilon_{\theta}} \| \epsilon  - \epsilon_{\theta}(y_t,t) \|^2_2 \\
&+ \lambda_{xt} \nabla_{\hat{y}_{0|t}^{'}} \| y_0  - \hat{y}_{0|t}^{'} \|^2_2 \\
&+ \lambda_{kt} \nabla_{\Phi_{\theta}} \| y_{acs} - y_{acs}^{'} \|^2_2
\end{aligned}$

\Until{converged} 
\end{algorithmic} 
\end{algorithm}

\begin{algorithm}
\caption{Sampling Process of dDiMo}\label{alg:sample}
\begin{algorithmic}[1]
\renewcommand{\algorithmicrequire}{\textbf{Input:}}
\Require $ y_T \sim \mathcal{N}\left(\mathbf{0}, \mathbf{I}\right)$, undersampling mask $\mathcal{M}$, partial scanned k-space $\tilde{y}$.
\For{$t=T,\dots,1$}
\State $y_{t} \leftarrow \mathcal{M}\left(\lambda_{t}\tilde{y} + \left(1-\lambda_{t}\right)y_{t}\right) + \left(1-\mathcal{M}\right)y_{t}$  
\hfill$\triangleright$ DC
\State $\hat{y}_{0|t} = \frac{1}{\sqrt{\bar{\alpha}_t}}\left(y_t - \sqrt{1-\bar{\alpha}_t}\epsilon_{\theta}\left(y_t, t\right)\right)$
\State $\hat{y}_{0|t}^{'} = \mathcal{F}\Psi_{\theta}\left(\mathcal{F}^{H}\hat{y}_{0|t}, t\right)$
\hfill$\triangleright$ $x$-$t$ prior
\State $\hat{y}_{0|t}^{''} = \Phi_{\theta}\left(\hat{y}_{0|t}^{'}\right)$
\hfill$\triangleright$ $k$-$t$ prior
\State $\hat{y}_{0|t}^{'''} = \mathbf{CG}\left(\hat{y}_{0|t}^{''}\right)$
\hfill$\triangleright$ CG
\State $z_t \sim \mathcal{N}\left(\mathbf{0}, \mathbf{I}\right)$ if $t>0$, else $z_t=0$
\State $
y_{t-1} = 
\frac{\sqrt{\bar{\alpha}_t}\left(1-\bar{\alpha}_{t-1}\right)}{1-\bar{\alpha}_t}y_t 
+ \frac{\sqrt{\bar{\alpha}_{t-1}}\beta_t}{1-\bar{\alpha}_t}\hat{y}_{0|t}^{'''} 
+ \sigma_{t}z_{t}$
\EndFor
\renewcommand{\algorithmicensure}{\textbf{Output:}}
\Ensure $y_0$ 
\end{algorithmic} 
\end{algorithm}

\subsection{Non-Cartesian Reconstruction}

\begin{figure*}
	\centering
	\includegraphics[width=\textwidth]{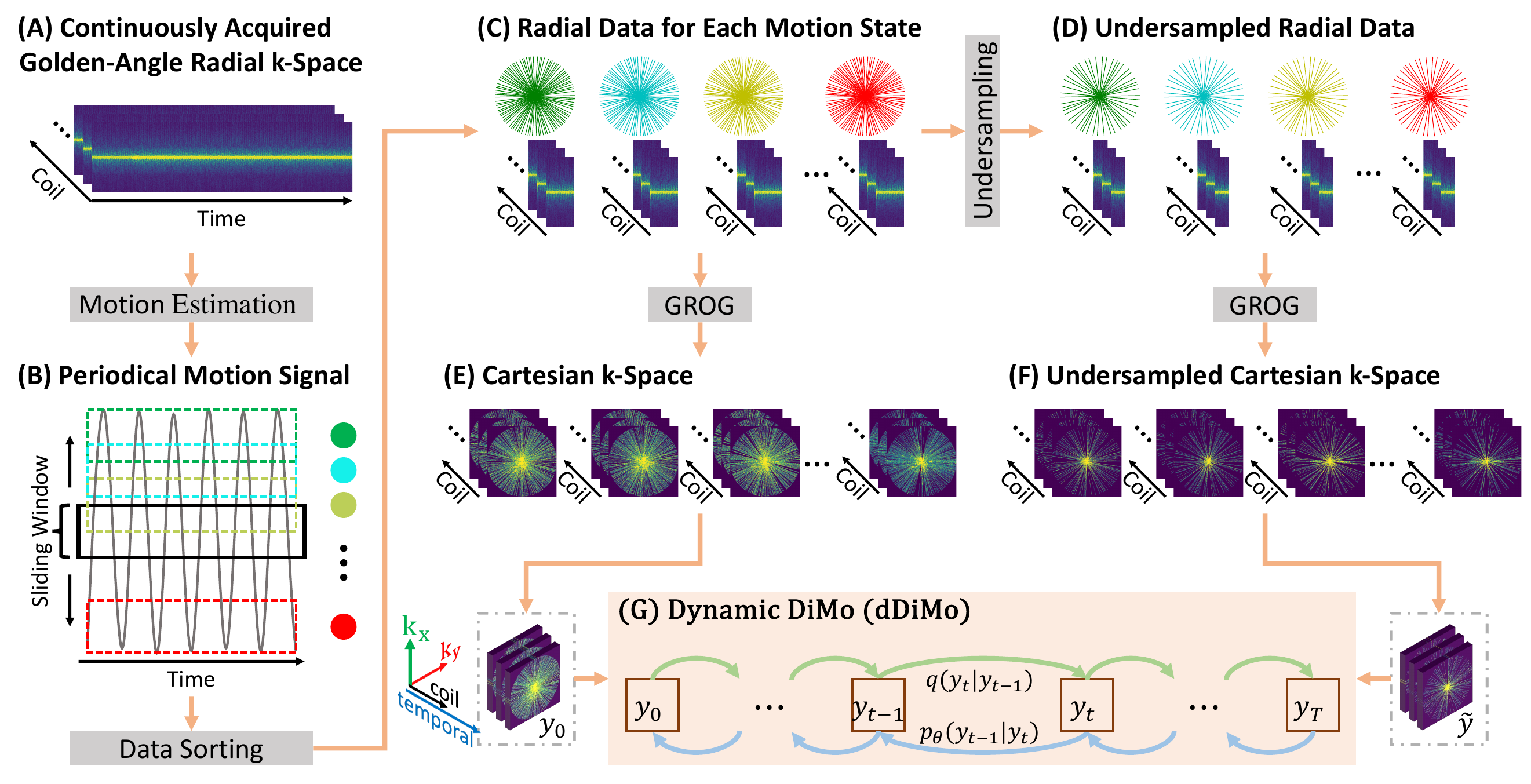}
	\caption{
    The overall pipeline of the proposed dDiMo framework for non-Cartesian radial dynamic MRI reconstruction.
    (A) Example of continuously acquired multi-coil golden-angle radial k-space data. (B) Respiratory motion signals estimated from (a) are used for data sorting and binning into motion states. (C) Radial k-space data corresponding to each motion state, randomly selected during training, serves as the reference data. (D) Undersampled radial k-space data is generated within each motion bin of (C) using random undersampling patterns. (E-F) The radial data pairs are gridded to Cartesian k-space using the GROG algorithm. (G) The gridded data pairs are fed to the dDiMo framework for reconstruction.
}
    \label{fig:dDiMo-radial}
\end{figure*}

While Figure~\ref{fig:dDiMo} illustrates the schematic diagram of dDiMo for reconstructing dynamic Caterisan k-space data, this proposed method can inherently be applied to non-Caterisian reconstruction as well. In this study, we demonstrate how dDiMo was adapted to reconstruct radial data acquired using a continuous stack-of-stars golden-angle radial acquisition scheme \cite{winkelmann2006optimal}. The updated framework consists of not only the diffusion modeling of dDiMo but also a few important preparation steps to adapt dDiMo with the non-Catesion k-space trajectory, as shown in Fig.~\ref{fig:dDiMo-radial}. We implemented a k-space binning, sorting, and data augmentation strategy to fully leverage the unique feature of this acquisition scheme for diffusion modeling. More specifically, in the data binning step, the motion signal was first estimated using a modified projection-based approach \cite{feng2016xd, spincemaille2011z} for the continuously acquired golden-angle radial k-space. The periodical motion signal was then divided into different motion states, and the data can be binned for each motion state. Instead of binning the data into a fixed number of non-overlapped bins, we implemented a \textit{flexible binning approach}, where a sliding window with a specific number of radial spokes can be used to bin at any motion state the continuously acquired data, allowing overlapped bins to capture the smooth transition between each motion state. This flexible binning approach behaves as a data augmentation when preparing the training data pairs for diffusion model training, with two advantages, first, it can generate extensive paired data to better utilize the continuous acquisition, and second, due to the bin overlap, it allows the network to see and then learn continuous motion state between the adjacent and overlapped bins, providing better generalizability than fixed binning at fixed motion state. 

After the binning using the sliding window, radial data within each bin can be sorted into the k-space to form reference training data for this specific motion state. Inspired by our previous work using sampling augmentation to improve learning performance \cite{liu2019santis}, the undersampled k-space data for each motion state were generated by randomly selecting a subset of radial spokes within each reference motion bin at each training iteration. This random sampling augmentation further diversifies the training data pairs, enabling the network to capture the complete features and patterns characteristic of radial data. The non-Cartesian radial k-space for reference and undersampled data was then mapped onto a Cartesian grid using GRAPPA operator gridding (GROG) \cite{seiberlich2007non}. We adopt self-calibrating GROG \cite{seiberlich2008self, benkert2018optimization} to shift radial data points to their nearest Cartesian grid locations, eliminating the need for separate calibration datasets. Reformatting the data into a Cartesian grid facilitates the efficient generation of binary Cartesian masks within the dDiMo framework, as shown in Figure~\ref{fig:dDiMo}, which can be directly used for reconstructing non-Caterisian radial data, as shown in Figure~\ref{fig:dDiMo-radial}(G).

\section{Experiments}
\subsection{Cartesian Cardiac Cine}
We evaluated dDiMo on 2D cardiac cine data from the CMRxRecon dataset \cite{wang2024cmrxrecon}. This cardiac cine data was acquired using a Cartesian TrueFISP sequence for imaging short-axis (SAX) and long-axis (LAX) heart views. The cardiac cycle was segmented into 12 to 25 phases with a temporal resolution of 50 ms. The scan parameters were as follows: the spatial resolution was 2.0$\times$2.0 mm$^2$, the slice thickness was 8.0 mm, and the slice gap was 4.0 mm. For each subject, 5 to 10 slices were acquired for the short-axis view, and three single slices were collected for the long-axis view at multi-view for imaging two, three, and four heart chambers.  A total number of 120 subjects were included in this study, including 96 (14364 slices) for training, 12 (1464 slices) for validation, and 12 (1560 slices) for testing. The raw multi-coil k-space data were compressed into 10 virtual coil elements using the method \cite{zhang2013coil} to reduce the computation memory cost. We tested multiple acceleration factors with 4$\times$, 8$\times$, and 10$\times$, and the center ACS lines covered 8\%, 4\%, and 3.2\% of the total phase-encoding lines, respectively, using a 1D variable-density undersampling scheme \cite{lustig2007sparse}. Network training was performed in 3 consecutive phases to capture temporal information in dDiMo, with both $\lambda_{xt}$ and $\lambda_{kt}$ set to 1 and $\lambda_{td}$ set to 0.015. The network parameters were initialized using Xavier initialization \cite{glorot2010understanding} and updated using the AdamW optimizer \cite{loshchilov2017decoupled} for 100 epochs at a learning rate of 0.0001 and total diffusion steps of 1000 with the ``cosine" schedule to ensure good model performance\cite{dhariwal2021diffusion}. We compared dDiMo to several reconstruction methods, including zero-filled image, a non-deep learning technique L+S \cite{otazo2015low} leveraging low-rank and sparsity for dynamic reconstruction, a non-diffusion deep learning method CRNN \cite{qin2018convolutional} for reconstructing dynamic data, and a diffusion method DiMo \cite{bian2024diffusion}, that treats each time frame independently without incorporating temporal information. The reconstruction settings for those competitive methods were recommended in their original papers along with available code.

\subsection{Dynamic Radial Lung Imaging}
We evaluated dDiMo on free-breathing volumetric lung MRI datasets acquired using a stack-of-stars golden-angle radial imaging sequence. MRI scans were conducted under a protocol approved by our institution’s institutional review board. Data acquisition was performed on a 3T MRI scanner (Siemens Healthcare, Erlangen, Germany) with the following imaging parameters: the flip angle was 15\textdegree, the echo time was 1.0 ms, the repetition time was 3.4 ms, the field of view of 410$\times$410 mm$^2$ and the voxel size of 2.1$\times$2.1$\times$2.5 mm$^3$. A total of 72 slices were acquired in coronal orientation, and 1700 radial spokes were acquired in each slice. The total scan time was 7 minutes. The data were acquired with an 18-channel coil and were subsequently compressed to 10 virtual coil elements\cite{zhang2013coil} to reduce the computation memory cost. Six datasets were included in this study, with 4 datasets being used for training/validation and the remaining 2 datasets for testing. Due to our data augmentation strategy, 48960 effective slice pairs were used during the training. We used a sliding window with 283 radial spokes to form each reference motion phase/bin and evaluated undersampling with 70, 35, and 17 spokes in each phase/bin, respectively. Network training was conducted in 6 phases to capture temporal information in dDiMo, with $\lambda_{xt}$ set to 1, $\lambda_{kt}$ set to 0.001, and $\lambda_{td}$ set to 0.015. Other network training settings are the same as those for Cartesian cardiac cine described above. For testing, 6 respiratory motion phases covering the entire respiratory cycle were used for evaluation. The results were compared with XD-GRASP\cite{feng2016xd} , a state-of-the-art method that leverages temporal sparsity, parallel imaging, and compressed sensing for reconstructing dynamic radial data, with reconstruction parameters recommended by the original paper. 

Results were quantitatively compared in terms of peak signal-to-noise ratio (PSNR), structure similarity (SSIM), normalized mean squared error (NMSE), and Tenengrad \cite{krotkov1989active} to evaluate noise performance, overall image recovery, overall reconstruction error, and image sharpness, respectively. All the programming in this study was implemented using Python language and the PyTorch package. All experiments were conducted on one NVIDIA A100 80GB GPU and an Intel Xeon 6338 CPU at Centos Linux system.

\subsection{Ablation Study}
In addition, ablation studies were conducted to evaluate several key aspects of our algorithm on the performance of reconstructing dynamic data, including investigating 1) the effect of the weighting factors $\lambda_{xt}$ and $\lambda_{kt}$ in the loss function (Eq.~\ref{eq:loss_function}) on reconstruction quality, illustrating the importance of inclusion of $x$-$t$ and $k$-$t$ regularization in the process; 2) the effect of the total number of diffusion time steps on reconstruction fidelity and its associated computational runtime; 3) the contribution of the nonlinear CG algorithm in the reverse diffusion process.

\section{Results}
\begin{figure*}
	\centering
	\includegraphics[width=\textwidth]{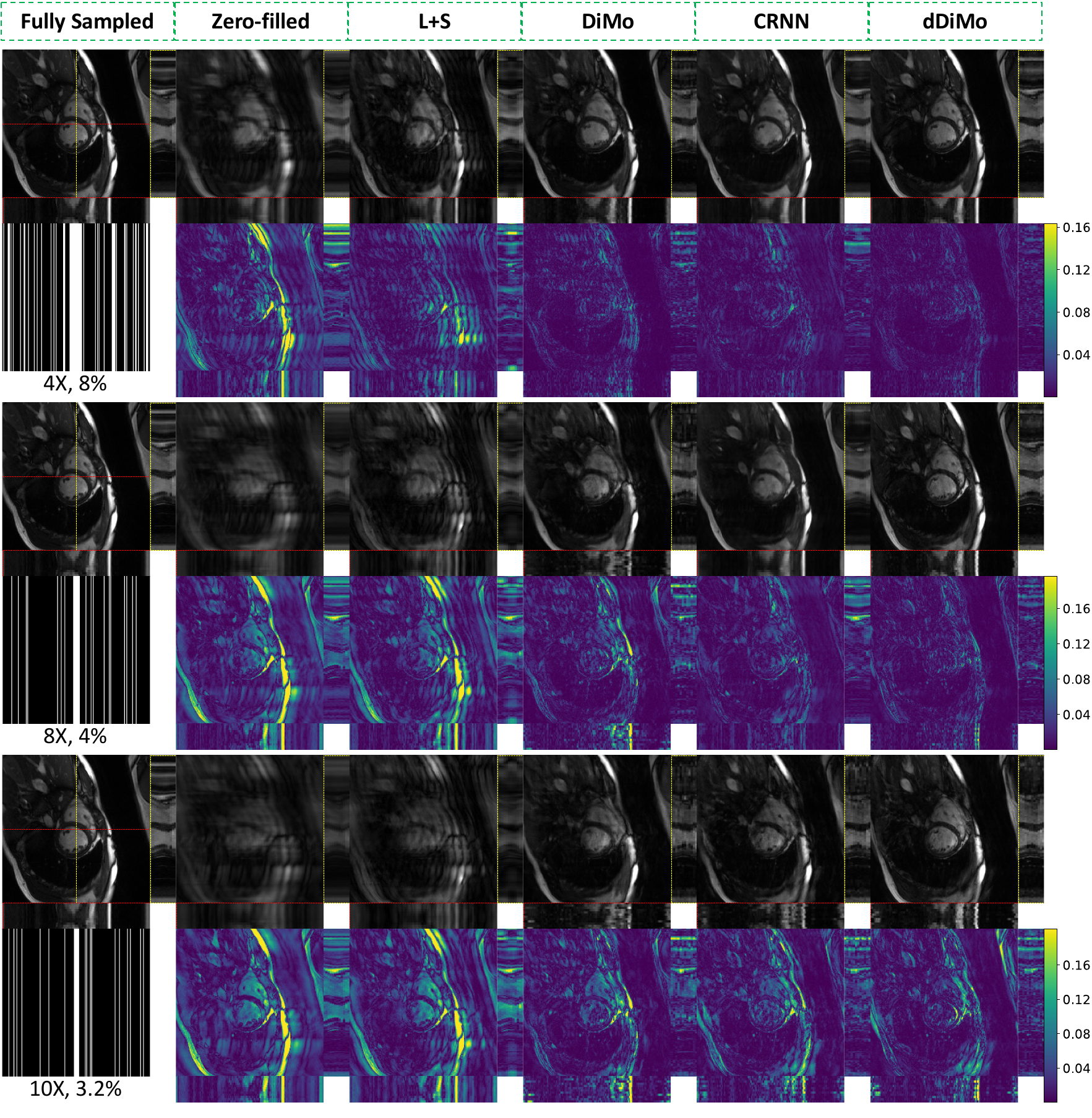}
	\caption{
        Qualitative comparison of different methods in spatial and spatiotemporal dimensions, along with corresponding error maps, for a cardiac cine in the short-axis view. Results are shown for undersampling rates of 4$\times$ (top), 8$\times$ (middle), and 16$\times$ (bottom). Spatiotemporal profiles along the yellow and red dotted lines are highlighted within yellow and red rectangles. The proposed method demonstrates superior performance in recovering fine structural details and preserving temporal coherence, even under highly undersampled conditions.
    }
    \label{fig:cine-sax}
\end{figure*}

\begin{figure*}
	\centering
	\includegraphics[width=\textwidth]{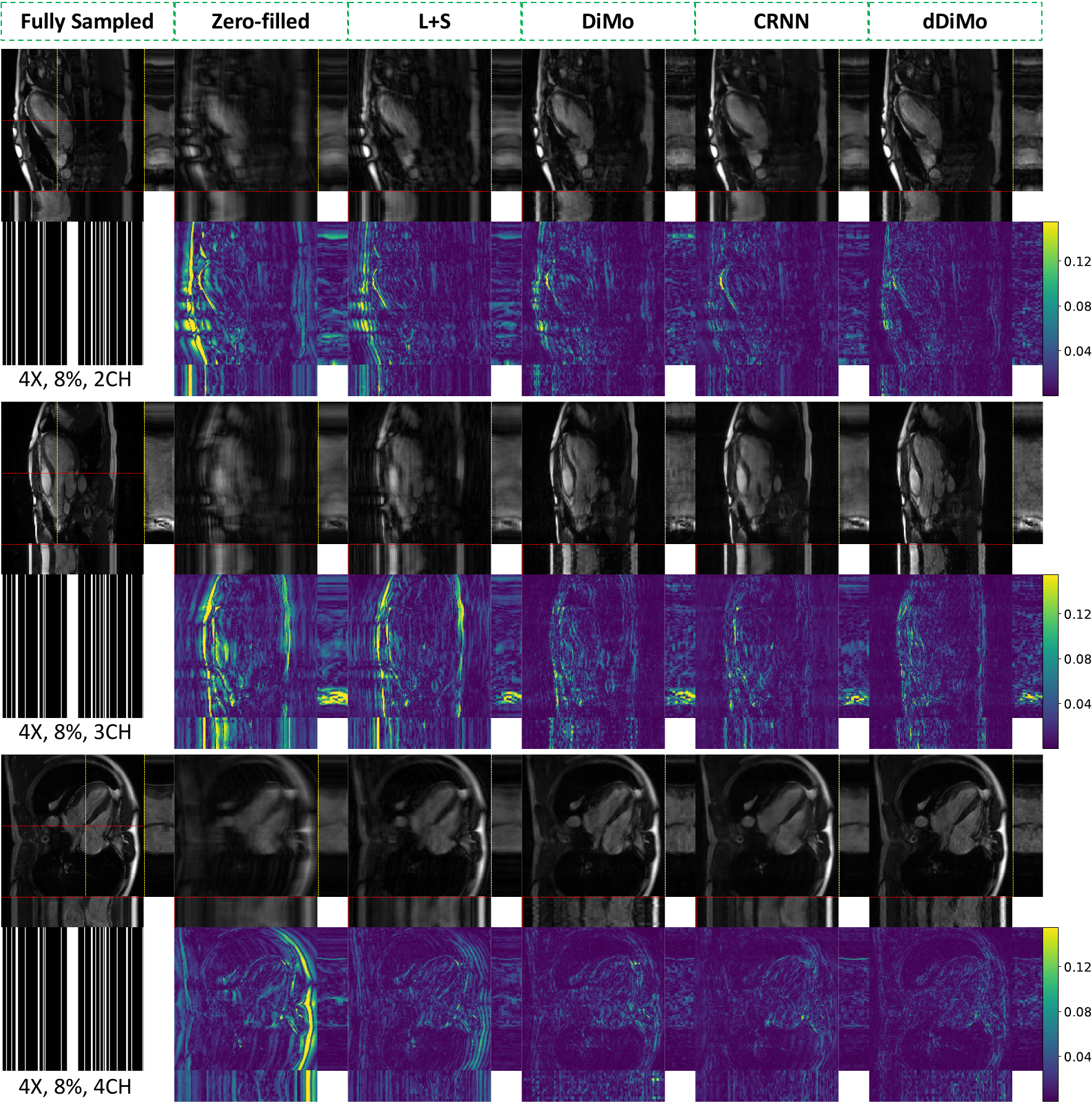}
	\caption{
        Qualitative comparison of different reconstruction methods in spatial and spatiotemporal dimensions, accompanied by corresponding error maps, for cardiac cine in long-axis views: two-chamber (2CH), three-chamber (3CH), and four-chamber (4CH). Spatiotemporal profiles along the yellow and red dotted lines are highlighted within yellow and red rectangles. Results are presented for an undersampling rate of 4$\times$. Additional examples at undersampling rates of 8$\times$ and 10$\times$ are provided in Supporting Information Figure S1 and Figure S2, respectively. The proposed method demonstrates superior performance in recovering fine spatial details and preserving temporal dynamics, even under challenging undersampling conditions.
    }
    \label{fig:cine-lax-4X}
\end{figure*}

\begin{figure*}
    \centering
    \includegraphics[width=\textwidth]{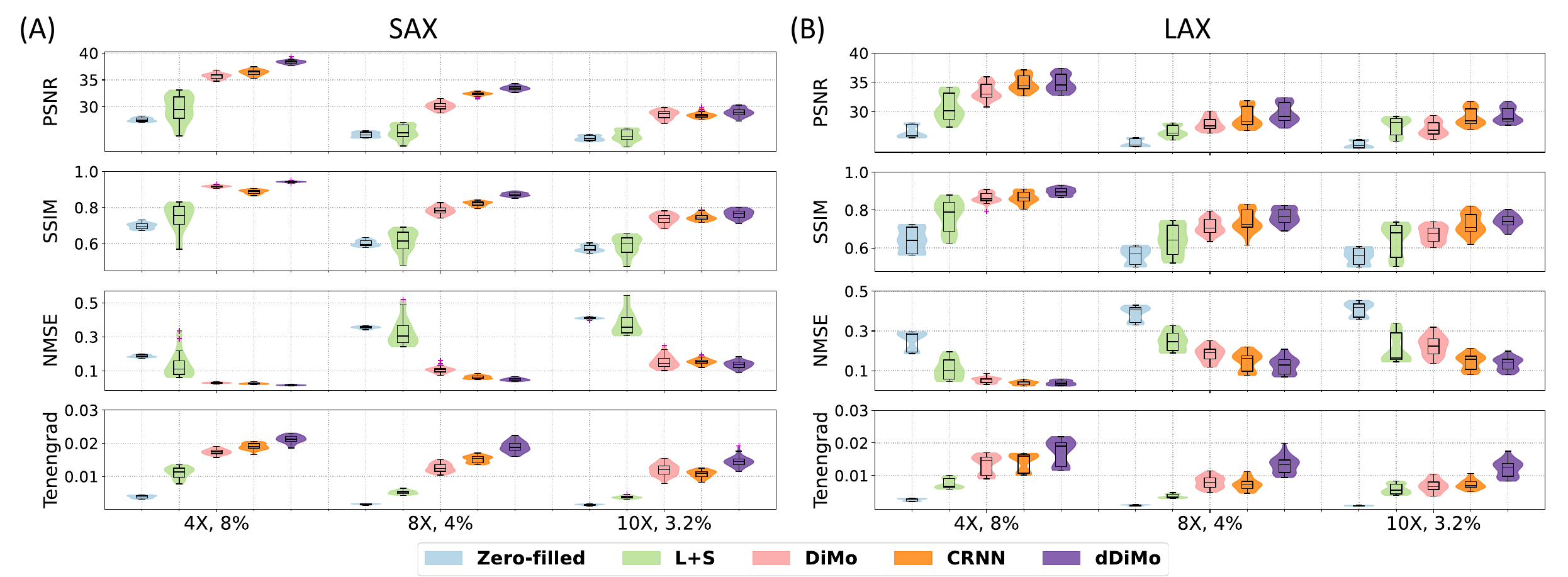}
    \caption{
    Violin plots of reconstruction metrics (PSNR, SSIM, and NMSE) for different methods (as indicated by the color-coded legend) at acceleration factors of 4$\times$, 8$\times$, and 10$\times$ for cardiac cine data. (A) Results for the short-axis (SAX) view. (B) Results for the long-axis (LAX) view. The plots highlight performance differences among methods, with the distributions illustrating reconstruction quality and stability across varying views and acceleration levels.
    }
    \label{fig:violin-sax-lax}
\end{figure*}

\begin{table*}
\caption{
Quantitative results (mean $\pm$ std) for accelerated cardiac cine SAX and LAX views at acceleration factors of 4$\times$, 8$\times$, and 10$\times$, with ACS lines covering 8\%, 4\%, and 3.2\% of the total phase-encoding lines. The best values for each metric are highlighted in bold.
}
\label{tbl:cine}
\centering
\begin{tabular}{l | l | l | cccc}
  \toprule
  View & Acc.  & Model   & PSNR $\uparrow$   & SSIM $\uparrow$    & NMSE $\downarrow$ & Tenengrad $\uparrow$        \\
  \midrule
  \multirow{15}{*}{SAX}
  & \multirow{5}{*}{4$\times$, 8\%}
  & Zero-filled   & 30.80 $\pm$ 0.30 & 0.8223 $\pm$ 0.0070 & 0.1766 $\pm$ 0.0053 & 0.0060 $\pm$ 0.0005      \\
  & & L+S           & 32.45 $\pm$ 2.48 & 0.8368 $\pm$ 0.0487 & 0.1391 $\pm$ 0.0812 & 0.0145 $\pm$ 0.0024      \\
  & & DiMo          & 38.77 $\pm$ 0.41 & 0.9491 $\pm$ 0.0023 & 0.0283 $\pm$ 0.0024 & 0.0227 $\pm$ 0.0009      \\
  & & CRNN          & 39.43 $\pm$ 0.50 & 0.9387 $\pm$ 0.0060 & 0.0245 $\pm$ 0.0038 & 0.0252 $\pm$ 0.0011 \\
  & & dDiMo         & \textbf{41.41 $\pm$ 0.27} & \textbf{0.9682 $\pm$ 0.0014} & \textbf{0.0154 $\pm$ 0.0017} & \textbf{0.0273 $\pm$ 0.0012}      \\
  \cmidrule{2-7}
  & \multirow{5}{*}{8$\times$, 4\%}
  & Zero-filled   & 27.92 $\pm$ 0.39 & 0.7569 $\pm$ 0.0076 & 0.3429 $\pm$ 0.0069 & 0.0026 $\pm$ 0.0002      \\
  & & L+S           & 28.27 $\pm$ 1.25 & 0.7526 $\pm$ 0.0420 & 0.3250 $\pm$ 0.0811 & 0.0069 $\pm$ 0.0007      \\
  & & DiMo          & 33.17 $\pm$ 0.54 & 0.8723 $\pm$ 0.0088 & 0.1031 $\pm$ 0.0139 & 0.0167 $\pm$ 0.0014     \\
  & & CRNN          & 35.24 $\pm$ 0.25 & 0.8994 $\pm$ 0.0060 & 0.0639 $\pm$ 0.0072 & 0.0203 $\pm$ 0.0011 \\
  & & dDiMo         & \textbf{36.46 $\pm$ 0.36} & \textbf{0.9270 $\pm$ 0.0048} & \textbf{0.0484 $\pm$ 0.0066} & \textbf{0.0244 $\pm$ 0.0017}    \\
  \cmidrule{2-7}
  & \multirow{5}{*}{10$\times$, 3.2\%}
  & Zero-filled   & 27.25 $\pm$ 0.38 & 0.7339 $\pm$ 0.0082 & 0.3995 $\pm$ 0.0042 & 0.0023 $\pm$ 0.0002     \\
  & & L+S           & 27.58 $\pm$ 1.07 & 0.7348 $\pm$ 0.0388 & 0.3770 $\pm$ 0.0747 & 0.0049 $\pm$ 0.0005     \\
  & & DiMo          & 31.57 $\pm$ 0.64 & 0.8394 $\pm$ 0.0126 & 0.1504 $\pm$ 0.0284 & 0.0156 $\pm$ 0.0018     \\
  & & CRNN          & 31.44 $\pm$ 0.49 & 0.8476 $\pm$ 0.0088 & 0.1529 $\pm$ 0.0147 & 0.0141 $\pm$ 0.0013 \\
  & & dDiMo         & \textbf{32.01 $\pm$ 0.64} & \textbf{0.8680 $\pm$ 0.0104} & \textbf{0.1351 $\pm$ 0.0190} & \textbf{0.0188 $\pm$ 0.0017}     \\
  \midrule
  \multirow{15}{*}{LAX}
  & \multirow{5}{*}{4$\times$, 8\%}
  & Zero-filled   & 29.98 $\pm$ 0.94 & 0.7935 $\pm$ 0.0231 & 0.2535 $\pm$ 0.0477 & 0.0036 $\pm$ 0.0004      \\
  & & L+S           & 33.75 $\pm$ 1.90 & 0.8648 $\pm$ 0.0296 & 0.1150 $\pm$ 0.0475 & 0.0095 $\pm$ 0.0019      \\
  & & DiMo          & 36.27 $\pm$ 1.17 & 0.9128 $\pm$ 0.0198 & 0.0601 $\pm$ 0.0144 & 0.0163 $\pm$ 0.0021      \\
  & & CRNN          & 37.61 $\pm$ 1.24 & 0.9163 $\pm$ 0.0177 & 0.0443 $\pm$ 0.0111 & 0.0174 $\pm$ 0.0022      \\
  & & dDiMo         & \textbf{37.82 $\pm$ 1.06} & \textbf{0.9393 $\pm$ 0.0095} & \textbf{0.0420 $\pm$ 0.0098} & \textbf{0.0209 $\pm$ 0.0031}      \\
  \cmidrule{2-7}
  & \multirow{5}{*}{8$\times$, 4\%}
  & Zero-filled   & 28.06 $\pm$ 0.61 & 0.7407 $\pm$ 0.0113 & 0.3882 $\pm$ 0.0410 & 0.0014 $\pm$ 0.0002      \\
  & & L+S           & 29.96 $\pm$ 0.73 & 0.7872 $\pm$ 0.0236 & 0.2534 $\pm$ 0.0449 & 0.0047 $\pm$ 0.0008      \\
  & & DiMo          & 31.12 $\pm$ 1.01 & 0.8270 $\pm$ 0.0295 & 0.1952 $\pm$ 0.0388 & 0.0102 $\pm$ 0.0015      \\
  & & CRNN & 32.19 $\pm$ 1.50 & 0.8438 $\pm$ 0.0354 & 0.1568 $\pm$ 0.0464 & 0.0093 $\pm$ 0.0018 \\
  & & dDiMo         & \textbf{32.69 $\pm$ 1.31} & \textbf{0.8603 $\pm$ 0.0222} & \textbf{0.1390 $\pm$ 0.0385} & \textbf{0.0159 $\pm$ 0.0020}      \\
  \cmidrule{2-7}
  & \multirow{5}{*}{10$\times$, 3.2\%}
  & Zero-filled   & 27.81 $\pm$ 0.58 & 0.7342 $\pm$ 0.0120 & 0.4112 $\pm$ 0.0383 & 0.0012 $\pm$ 0.0002      \\
  & & L+S           & 30.63 $\pm$ 0.92 & 0.7860 $\pm$ 0.0205 & 0.2214 $\pm$ 0.0622 & 0.0071 $\pm$ 0.0019      \\
  & & DiMo          & 30.24 $\pm$ 1.04 & 0.8031 $\pm$ 0.0306 & 0.2392 $\pm$ 0.0502 & 0.0087 $\pm$ 0.0019      \\
  & & CRNN & 32.13 $\pm$ 1.24 & 0.8306 $\pm$ 0.0287 & 0.1567 $\pm$ 0.0397 & 0.0091 $\pm$ 0.0014 \\
  & & dDiMo         & \textbf{32.45 $\pm$ 1.13} & \textbf{0.8462 $\pm$ 0.0222} & \textbf{0.1449 $\pm$ 0.0337} & \textbf{0.0146 $\pm$ 0.0017}      \\
  \bottomrule
\end{tabular}
\end{table*}

\subsection{Results of Cardiac Cine}
Exemplary reconstructions from various methods for cine images at different acceleration factors and cardiac views are shown in Figure~\ref{fig:cine-sax}, Figure~\ref{fig:cine-lax-4X}, and Supporting Information Figure S1 and Figure S2. The proposed dDiMo method demonstrates superior performance in artifact suppression, detail preservation, and accurate cardiac motion capture, producing sharp and clear reconstructions. In contrast, competing methods, including L+S, DiMo, and CRNN, exhibit notable artifacts, spatial blurring, and temporal misalignment, particularly at higher acceleration factors. Exemplary $x$-$t$ profiles extracted along the vertical yellow dashed line and horizontal red dashed line, as indicated by the yellow and red dashed rectangles on the spatial reconstruction images, are presented to illustrate temporal alignment across cardiac frames. Reconstructions from dDiMo achieve better alignment along the cardiac phase, with clearer boundary definitions that closely resemble the fully sampled reference image compared to other methods. These observations are further supported by error maps of the reconstructed spatial images and $x$-$t$ profiles. Zero-filled reconstructions show the largest errors, followed by L+S, DiMo, and CRNN. In general, CRNN shows improved performance over DiMo by leveraging temporal characterization, while DiMo, which processes each frame independently, results in misaligned cardiac phases due to the lack of temporal information integration during the diffusion process.

Quantitative results for SAX and LAX views of cine reconstruction are summarized in Table~\ref{tbl:cine}. Metrics such as PSNR, SSIM, NMSE, and Tenengrad confirm that dDiMo consistently outperforms baseline methods in acceleration factors (4$\times$, 8$\times$, and 10$\times$) and cardiac views. Even at higher acceleration rates, where reconstruction becomes more challenging, dDiMo demonstrates superior performance across all metrics, highlighting its robustness. Violin plots in Figure~\ref{fig:violin-sax-lax} provide a visual comparison of the distributions of PSNR, SSIM, NMSE, and Tenengrad metrics for different acceleration settings. dDiMo achieves the highest median values for PSNR, SSIM, and Tenengrad and the lowest NMSE, with minimal variability and fewer outliers. This demonstrates its ability to produce stable and reliable reconstructions under challenging conditions. These quantitative results validate the qualitative findings and confirm the effectiveness of dDiMo for Cartesian-acquired dynamic MRI reconstruction.

\subsection{Results of Radial Lung Imaging}
\begin{figure*}
    \centering
    \includegraphics[width=\textwidth]{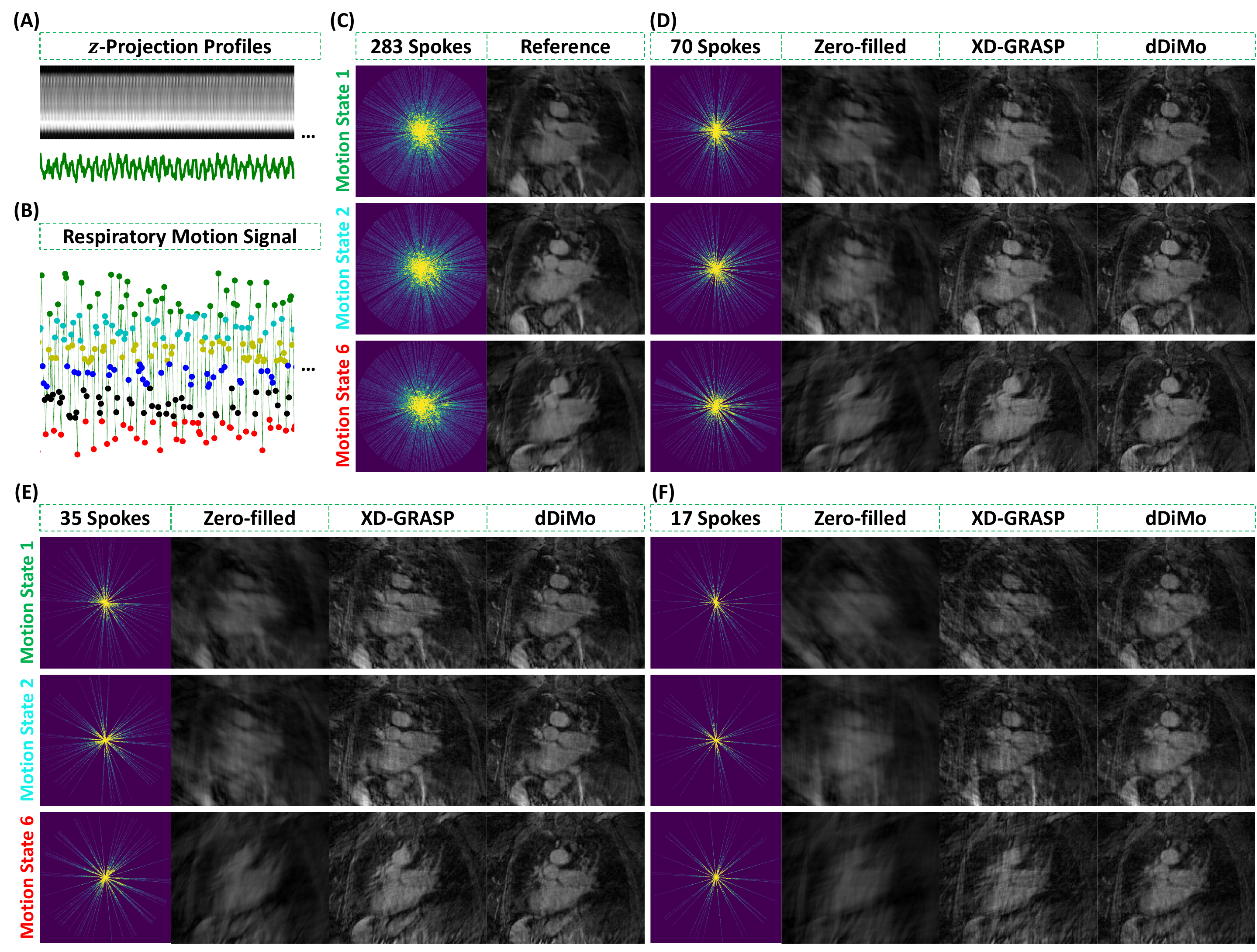}
    \caption{
        Qualitative comparison of reconstructions from continuously acquired golden-angle radial multicoil free-breathing lung data using Zero-filled, XD-GRASP, and dDiMo methods with varying numbers of spokes per motion state. Representative results are shown for Motion States 1, 2, and 6, illustrating the ability of each method to handle undersampled dynamic lung imaging.
        (A) Processed respiratory motion signals (below) extracted from z-projection profiles (above).
        (B) Radial k-space data continuously acquired and sorted into respiratory states based on motion signals extracted directly from the data. Different colors represent distinct motion states, with an equal number of spokes grouped in each state.
        (C) Reference image reconstructed using 283 spokes.
        (D) Images reconstructed from each method using 70 spokes.
        (E) Images reconstructed from each method using 35 spokes.
        (F) Images reconstructed from each method using 17 spokes.
        The proposed dDiMo method demonstrates superior performance by effectively suppressing artifacts, recovering fine structural details, and maintaining motion consistency across states.
    }
    \label{fig:lung}
\end{figure*}

\begin{table*}
\caption{
Quantitative results (mean $\pm$ std) for continuously acquired golden-angle radial multicoil free-breathing lung data with 70, 35, and 17 spokes. The best values for each metric are highlighted in bold.
}
\label{tbl:lung}
\centering
\begin{tabular}{l | l | cccc}
  \toprule
  Acc.  & Model   & PSNR $\uparrow$   & SSIM $\uparrow$    & NMSE $\downarrow$  & Tenengrad $\uparrow$        \\
  \midrule
  \multirow{3}{*}{70 spokes}
  & Zero-filled & 31.69 $\pm$ 1.15 & 0.8499 $\pm$ 0.0165 & 0.1236 $\pm$ 0.0208 & 0.0078 $\pm$ 0.0017      \\
  & XD-GRASP    & 32.08 $\pm$ 1.09 & 0.8475 $\pm$ 0.0196 & 0.1142 $\pm$ 0.0258 & \textbf{0.0309 $\pm$ 0.0060}      \\
  & dDiMo       & \textbf{33.41 $\pm$ 1.01} & \textbf{0.8675 $\pm$ 0.0145} & \textbf{0.0834 $\pm$ 0.0157} & 0.0286 $\pm$ 0.0050      \\
  \midrule
  \multirow{3}{*}{35 spokes}
  & Zero-filled & 29.07 $\pm$ 1.19 & 0.7687 $\pm$ 0.0267 & 0.2254 $\pm$ 0.0355 & 0.0042 $\pm$ 0.0010      \\
  & XD-GRASP    & 31.46 $\pm$ 1.13 & 0.8005 $\pm$ 0.0189 & 0.1296 $\pm$ 0.0188 & \textbf{0.0263 $\pm$ 0.0051}      \\
  & dDiMo       & \textbf{32.45 $\pm$ 1.35} & \textbf{0.8406 $\pm$ 0.0190} & \textbf{0.1039 $\pm$ 0.0193} & 0.0189 $\pm$ 0.0052      \\
  \midrule
  \multirow{3}{*}{17 spokes}
  & Zero-filled & 27.62 $\pm$ 1.14 & 0.7124 $\pm$ 0.0195 & 0.3136 $\pm$ 0.0453 & 0.0022 $\pm$ 0.0005      \\
  & XD-GRASP    & 30.62 $\pm$ 1.04 & 0.7667 $\pm$ 0.0179 & 0.1575 $\pm$ 0.0226 & \textbf{0.0197 $\pm$ 0.0039}      \\
  & dDiMo       & \textbf{30.63 $\pm$ 1.55} & \textbf{0.8013 $\pm$ 0.0242} & \textbf{0.1623 $\pm$ 0.0492} & 0.0113 $\pm$ 0.0039      \\
  \bottomrule
\end{tabular}
\end{table*}

Figure~\ref{fig:lung} presents a qualitative comparison of lung reconstructions with 70, 35, and 17 spokes across motion states using Zero-filled, XD-GRASP, and the proposed dDiMo method. Reconstructions are shown for representative motion states (Motion State 1, Motion State 2, and Motion State 6), highlighting the effectiveness of each method in handling undersampled dynamic lung imaging. Zero-filled reconstructions exhibit severe artifacts and significant loss of structural detail, particularly in motion-affected regions. For instance, Motion State 1 demonstrates extensive blurring and cannot delineate key anatomical features. XD-GRASP reduces artifacts and partially restores structural details compared to Zero-filled; however, residual blurring, missing subtle textures, and noise remain apparent, especially at higher undersampling levels (e.g., 17 spokes). In contrast, dDiMo achieves superior performance across all motion states by effectively suppressing artifacts, recovering fine structural details, and maintaining consistency across motion states. In the most challenging scenario, with 17 spokes, dDiMo reconstructs most of the structural details, producing sharper boundaries and enhanced structural integrity across respiratory phases. These qualitative results highlight the advantages of the proposed dDiMo method over traditional approaches, demonstrating its capacity to restore structural details and improve motion-resolved reconstructions.

\begin{figure}
    \centering
    \includegraphics[width=0.5\textwidth]{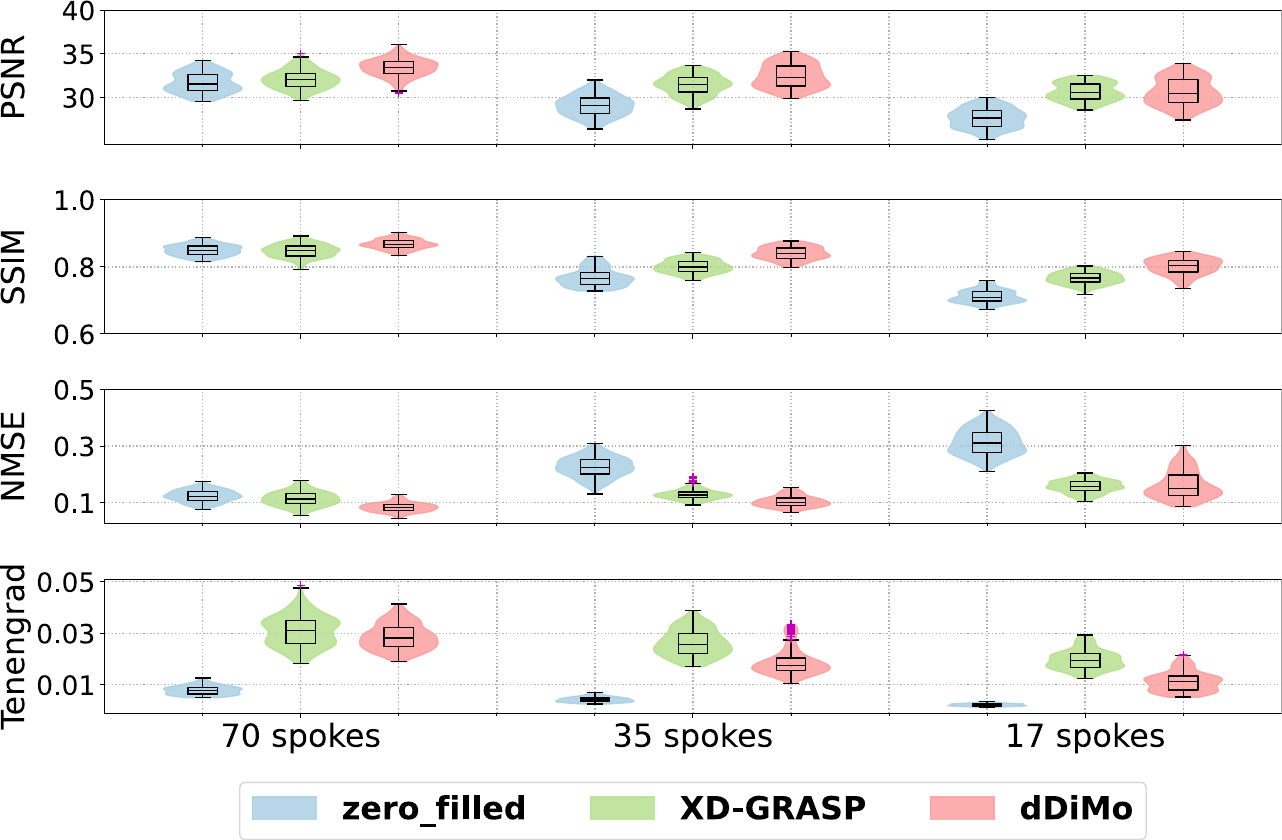}
    \caption{
        Violin plots of reconstruction metrics for different methods (Zero-filled, XD-GRASP, and dDiMo) across varying undersampling levels (70, 35, and 17 spokes) for continuously acquired golden-angle radial multicoil free-breathing lung data. The plots demonstrate the robust performance of dDiMo, with narrower distributions and fewer outliers, highlighting its ability to consistently achieve high-fidelity reconstructions compared to other methods.
    }
    \label{fig:violin-lung}
\end{figure}

Quantitative results comparing reconstruction methods at different undersampling levels (70, 35, and 17 spokes) are summarized in Table~\ref{tbl:lung}. dDiMo consistently achieves the highest PSNR and overall image similarity with respect to the reference across all undersampling levels, outperforming both Zero-filled and XD-GRASP. While XD-GRASP achieves the highest Tenengrad score at all undersampling levels, it performs worse in terms of overall image similarity. This is attributed to the presence of noise in XD-GRASP reconstructions, which leads to inflated Tenengrad scores, as also observed in Figure~\ref{fig:lung}. Violin plots in Figure~\ref{fig:violin-lung} provide a visual comparison of the distributions of PSNR, SSIM, NMSE, and Tenengrad metrics for different acceleration settings. dDiMo exhibits narrower distributions and fewer outliers compared to Zero-filled and XD-GRASP in most cases, reflecting its stable and robust performance.

\subsection{Results of the Ablation Study}
The ablation study evaluated the effectiveness of the proposed temporal guidance priors, including $x$-$t$ and $k$-$t$ priors, by analyzing results with different weighting factors $\lambda_{xt}$ and $\lambda_{kt}$ from Eq. \eqref{eq:loss_function} on cardiac cine data, as presented in Supporting Information Figure S3. The influence of the number of reverse diffusion steps on reconstruction fidelity and computational runtime was analyzed on cardiac cine data, as depicted in Supporting Information Figure S4. Additionally, the impact of incorporating the CG module during inference was assessed on dynamic lung data, with results shown in Supporting Information Figure S5. 

\section{Discussion}
This study highlights the potential of the dDiMo framework for reconstructing high-quality images from undersampled dynamic multi-coil Cartesian cardiac cine data as well as non-Cartesian golden-angle radial lung data. By integrating temporal information from time-resolved dimensions, dDiMo allows for the simultaneous recovery of spatial features within individual frames and the temporal dynamics across frames. 

Incorporating $x$-$t$ and $k$-$t$ regularization—derived directly from time-resolved data provides valuable guidance during the diffusion process. This approach ensures accurate temporal alignment and enhances the recovery of fine image details, as is shown by comparing dDiMo with other competitive methods with (e.g., CRNN) or without (e.g., DiMo) temporal characterization. Also, it was highlighted in the ablation study results (Supporting Information Figure S3) that the reconstruction performance of the $x$-$t$ prior improves significantly as its weighting factor $\lambda_{xt}$ increases, with optimal results achieved at $\lambda_{xt}=1$. A similar trend is observed for the $k$-$t$ prior, with the performance increasing when the weighting factors increase to 1. This illustrates the contribution of $x$-$t$ and $k$-$t$ regularization to improving reconstruction performance in the cardiac cine data. However, notably, the performance wasn't further improved with a higher $x$-$t$ and $k$-$t$ weight, indicating optimal performance exists with dataset-specific tunning of those numbers. While not shown in the result, we realized the result for the golden-angle radial lung data relies more on the $x$-$t$ component rather than the $k$-$t$ characterization, as we used a relatively small weight 0.001 for $k$-$t$ component. This is mainly because the learning of the accurate $k$-$t$ priors in the center ACS region of radial k-space is extremely challenging and prone to error due to the highly oversampled k-space center and small ACS coverage, especially at extremely few radial spokes (e.g., 35 or 17 spokes at Figure ~\ref{fig:lung}) \cite{feng2020grasp}. The CG layer also plays an important role in enhancing performance within this framework. Supporting Information Figure S4 illustrates that while dDiMo without CG can recover a substantial amount of image structure from undersampled data, it often results in less clear images. By implementing temporal sparsity through CG optimization, sharper reconstructions with finer details are achieved as the weighting factor $\lambda_{td}$ in Eq. \eqref{eq:cg_objective} increases. In the lung dataset, the optimal value for $\lambda_{td}$ was found to be 0.015. Higher values led to degraded reconstructions, which were characterized by noise and blurring. Nevertheless, we demonstrate the inclusion of $x$-$t$ and $k$-$t$ priors and CG into the diffusion modeling can be beneficial to characterize dynamic information in dynamic MRI reconstruction; care has to be cast to carefully leverage the good aspects of those modules. As with many dedicated deep learning algorithms with fine structures and components, hyperparameter tuning is never trivial and forgiven, including for dDiMo. Further research is warranted to explore robust, cost-effective, and hopefully automatic ways to facilitate this process \cite{bian2025multi}.

The diffusion modeling-based methods also have limitations. One notable limitation of dDiMo is its relatively slow inference speed, which stems from the sequential nature of the diffusion modeling process. Compared to other deep learning methods where reconstruction can be conducted by running the network once, the reverse diffusion requires the iterative operation of network inference for extensive steps, in our case, 1000 steps. Supporting Information Figure S5 highlights the trade-off between the number of reverse diffusion steps and reconstruction quality. Reconstructions with fewer than 100 steps exhibit residual noise while increasing the number of steps enhances detail recovery and noise suppression. In our study, 1000 reverse diffusion steps were employed, balancing reconstruction accuracy and computational efficiency. However, this leads to an average inference time of 6 minutes for reconstructing one image volume. To address this challenge, future research could leverage the potential of latent diffusion models, as highlighted in recent advancements \cite{rombach2022high}. By redefining the diffusion process in a compressed latent space rather than the pixel space, these models significantly reduce computational complexity while maintaining the ability to generate high-quality data restoration. Additionally, future work could explore integrating other temporal feature-aware and efficient data compression techniques, such as subspace learning \cite{jun2024zero, sandino2021deep}, into diffusion modeling to alleviate the current high computational burden.

\section{Conclusion}
This study presents a new diffusion modeling technique that integrates temporal information from time-resolved data to improve the diffusion process used for reconstructing undersampled dynamic MRI. The method demonstrates strong performance with both Cartesian and non-Cartesian data. The findings indicate that this innovative approach has the potential to deliver high-quality image reconstruction for various clinical applications.


\bibliographystyle{IEEEtran}
\bibliography{IEEEabrv,references}

\begin{thebibliography}{10}
\providecommand{\url}[1]{#1}
\csname url@samestyle\endcsname
\providecommand{\newblock}{\relax}
\providecommand{\bibinfo}[2]{#2}
\providecommand{\BIBentrySTDinterwordspacing}{\spaceskip=0pt\relax}
\providecommand{\BIBentryALTinterwordstretchfactor}{4}
\providecommand{\BIBentryALTinterwordspacing}{\spaceskip=\fontdimen2\font plus
\BIBentryALTinterwordstretchfactor\fontdimen3\font minus
  \fontdimen4\font\relax}
\providecommand{\BIBforeignlanguage}[2]{{%
\expandafter\ifx\csname l@#1\endcsname\relax
\typeout{** WARNING: IEEEtran.bst: No hyphenation pattern has been}%
\typeout{** loaded for the language `#1'. Using the pattern for}%
\typeout{** the default language instead.}%
\else
\language=\csname l@#1\endcsname
\fi
#2}}
\providecommand{\BIBdecl}{\relax}
\BIBdecl

\bibitem{sodickson1997simultaneous}
D.~K. Sodickson and W.~J. Manning, ``Simultaneous acquisition of spatial
  harmonics (smash): fast imaging with radiofrequency coil arrays,''
  \emph{Magnetic resonance in medicine}, vol.~38, no.~4, pp. 591--603, 1997.

\bibitem{pruessmann1999sense}
K.~P. Pruessmann, M.~Weiger, M.~B. Scheidegger, and P.~Boesiger, ``Sense:
  Sensitivity encoding for fast mri,'' \emph{Magnetic Resonance in Medicine},
  vol.~42, no.~5, pp. 952--962, 1999.

\bibitem{griswold2002grappa}
M.~A. Griswold, P.~M. Jakob, R.~M. Heidemann, M.~Nittka, V.~Jellus, J.~Wang,
  B.~Kiefer, and A.~Haase, ``Generalized autocalibrating partially parallel
  acquisitions (grappa),'' \emph{Magnetic Resonance in Medicine}, vol.~47,
  no.~6, pp. 1202--1210, 2002.

\bibitem{lustig2007sparse}
M.~Lustig, D.~Donoho, and J.~M. Pauly, ``Sparse mri: The application of
  compressed sensing for rapid mr imaging,'' \emph{Magnetic Resonance in
  Medicine: An Official Journal of the International Society for Magnetic
  Resonance in Medicine}, vol.~58, no.~6, pp. 1182--1195, 2007.

\bibitem{lingala2011accelerated}
S.~G. Lingala, Y.~Hu, E.~DiBella, and M.~Jacob, ``Accelerated dynamic mri
  exploiting sparsity and low-rank structure: kt slr,'' \emph{IEEE transactions
  on medical imaging}, vol.~30, no.~5, pp. 1042--1054, 2011.

\bibitem{christodoulou2013high}
A.~G. Christodoulou, H.~Zhang, B.~Zhao, T.~K. Hitchens, C.~Ho, and Z.-P. Liang,
  ``High-resolution cardiovascular mri by integrating parallel imaging with
  low-rank and sparse modeling,'' \emph{IEEE Transactions on Biomedical
  Engineering}, vol.~60, no.~11, pp. 3083--3092, 2013.

\bibitem{haldar2013low}
J.~P. Haldar, ``Low-rank modeling of local $ k $-space neighborhoods (loraks)
  for constrained mri,'' \emph{IEEE transactions on medical imaging}, vol.~33,
  no.~3, pp. 668--681, 2013.

\bibitem{dong2014compressive}
W.~Dong, G.~Shi, X.~Li, Y.~Ma, and F.~Huang, ``Compressive sensing via nonlocal
  low-rank regularization,'' \emph{IEEE transactions on image processing},
  vol.~23, no.~8, pp. 3618--3632, 2014.

\bibitem{shin2014calibrationless}
P.~J. Shin, P.~E. Larson, M.~A. Ohliger, M.~Elad, J.~M. Pauly, D.~B. Vigneron,
  and M.~Lustig, ``Calibrationless parallel imaging reconstruction based on
  structured low-rank matrix completion,'' \emph{Magnetic resonance in
  medicine}, vol.~72, no.~4, pp. 959--970, 2014.

\bibitem{otazo2015low}
R.~Otazo, E.~Candes, and D.~K. Sodickson, ``Low-rank plus sparse matrix
  decomposition for accelerated dynamic mri with separation of background and
  dynamic components,'' \emph{Magnetic resonance in medicine}, vol.~73, no.~3,
  pp. 1125--1136, 2015.

\bibitem{zhang2015accelerating}
T.~Zhang, J.~M. Pauly, and I.~R. Levesque, ``Accelerating parameter mapping
  with a locally low rank constraint,'' \emph{Magnetic resonance in medicine},
  vol.~73, no.~2, pp. 655--661, 2015.

\bibitem{he2016accelerated}
J.~He, Q.~Liu, A.~G. Christodoulou, C.~Ma, F.~Lam, and Z.-P. Liang,
  ``Accelerated high-dimensional mr imaging with sparse sampling using low-rank
  tensors,'' \emph{IEEE transactions on medical imaging}, vol.~35, no.~9, pp.
  2119--2129, 2016.

\bibitem{wang2016accelerating}
S.~Wang, Z.~Su, L.~Ying, X.~Peng, S.~Zhu, F.~Liang, D.~Feng, and D.~Liang,
  ``Accelerating magnetic resonance imaging via deep learning,'' \emph{2016
  IEEE 13th International Symposium on Biomedical Imaging (ISBI)}, pp.
  514--517, 2016.

\bibitem{sun2016deep}
Y.~Yang, J.~Sun, H.~Li, and Z.~Xu, ``Deep admm-net for compressive sensing
  mri,'' \emph{Advances in neural information processing systems}, vol.~29,
  2016.

\bibitem{yang2017dagan}
G.~Yang, S.~Yu, H.~Dong, G.~Slabaugh, P.~L. Dragotti, X.~Ye, F.~Liu,
  S.~Arridge, J.~Keegan, Y.~Guo \emph{et~al.}, ``Dagan: Deep de-aliasing
  generative adversarial networks for fast compressed sensing mri
  reconstruction,'' \emph{IEEE transactions on medical imaging}, vol.~37,
  no.~6, pp. 1310--1321, 2017.

\bibitem{aggarwal2018modl}
H.~K. Aggarwal, M.~P. Mani, and M.~Jacob, ``Modl: Model-based deep learning
  architecture for inverse problems,'' \emph{IEEE transactions on medical
  imaging}, vol.~38, no.~2, pp. 394--405, 2018.

\bibitem{hammernik2018learning}
K.~Hammernik, T.~Klatzer, E.~Kobler, M.~P. Recht, D.~K. Sodickson, T.~Pock, and
  F.~Knoll, ``Learning a variational network for reconstruction of accelerated
  mri data,'' \emph{Magnetic resonance in medicine}, vol.~79, no.~6, pp.
  3055--3071, 2018.

\bibitem{eo2018kiki}
T.~Eo, Y.~Jun, T.~Kim, J.~Jang, H.-J. Lee, and D.~Hwang, ``Kiki-net:
  cross-domain convolutional neural networks for reconstructing undersampled
  magnetic resonance images,'' \emph{Magnetic resonance in medicine}, vol.~80,
  no.~5, pp. 2188--2201, 2018.

\bibitem{zhu2018image}
B.~Zhu, J.~Z. Liu, S.~F. Cauley, B.~R. Rosen, and M.~S. Rosen, ``Image
  reconstruction by domain-transform manifold learning,'' \emph{Nature}, vol.
  555, no. 7697, pp. 487--492, 2018.

\bibitem{liu2019santis}
F.~Liu, A.~Samsonov, L.~Chen, R.~Kijowski, and L.~Feng, ``Santis:
  sampling-augmented neural network with incoherent structure for mr image
  reconstruction,'' \emph{Magnetic resonance in medicine}, vol.~82, no.~5, pp.
  1890--1904, 2019.

\bibitem{ahmad2020plug}
R.~Ahmad, C.~A. Bouman, G.~T. Buzzard, S.~Chan, S.~Liu, E.~T. Reehorst, and
  P.~Schniter, ``Plug-and-play methods for magnetic resonance imaging: Using
  denoisers for image recovery,'' \emph{IEEE signal processing magazine},
  vol.~37, no.~1, pp. 105--116, 2020.

\bibitem{akccakaya2019scan}
M.~Ak{\c{c}}akaya, S.~Moeller, S.~Weing{\"a}rtner, and K.~U{\u{g}}urbil,
  ``Scan-specific robust artificial-neural-networks for k-space interpolation
  (raki) reconstruction: Database-free deep learning for fast imaging,''
  \emph{Magnetic resonance in medicine}, vol.~81, no.~1, pp. 439--453, 2019.

\bibitem{sriram2020end}
A.~Sriram, J.~Zbontar, T.~Murrell, A.~Defazio, C.~L. Zitnick, N.~Yakubova,
  F.~Knoll, and P.~Johnson, ``End-to-end variational networks for accelerated
  mri reconstruction,'' \emph{Medical Image Computing and Computer Assisted
  Intervention--MICCAI 2020: 23rd International Conference, Lima, Peru, October
  4--8, 2020, Proceedings, Part II 23}, pp. 64--73, 2020.

\bibitem{schlemper2017deep}
J.~Schlemper, J.~Caballero, J.~V. Hajnal, A.~N. Price, and D.~Rueckert, ``A
  deep cascade of convolutional neural networks for dynamic mr image
  reconstruction,'' \emph{IEEE transactions on Medical Imaging}, vol.~37,
  no.~2, pp. 491--503, 2017.

\bibitem{qin2018convolutional}
C.~Qin, J.~Schlemper, J.~Caballero, A.~N. Price, J.~V. Hajnal, and D.~Rueckert,
  ``Convolutional recurrent neural networks for dynamic mr image
  reconstruction,'' \emph{IEEE transactions on medical imaging}, vol.~38,
  no.~1, pp. 280--290, 2018.

\bibitem{qin2021complementary}
C.~Qin, J.~Duan, K.~Hammernik, J.~Schlemper, T.~K{\"u}stner, R.~Botnar,
  C.~Prieto, A.~N. Price, J.~V. Hajnal, and D.~Rueckert, ``Complementary
  time-frequency domain networks for dynamic parallel mr image
  reconstruction,'' \emph{Magnetic Resonance in Medicine}, vol.~86, no.~6, pp.
  3274--3291, 2021.

\bibitem{kustner2020cinenet}
T.~K{\"u}stner, N.~Fuin, K.~Hammernik, A.~Bustin, H.~Qi, R.~Hajhosseiny, P.~G.
  Masci, R.~Neji, D.~Rueckert, R.~M. Botnar \emph{et~al.}, ``Cinenet: deep
  learning-based 3d cardiac cine mri reconstruction with multi-coil
  complex-valued 4d spatio-temporal convolutions,'' \emph{Scientific reports},
  vol.~10, no.~1, p. 13710, 2020.

\bibitem{dar2020prior}
S.~U. Dar, M.~Yurt, M.~Shahdloo, M.~E. Ild{\i}z, B.~T{\i}naz, and
  T.~{\c{C}}ukur, ``Prior-guided image reconstruction for accelerated
  multi-contrast mri via generative adversarial networks,'' \emph{IEEE Journal
  of Selected Topics in Signal Processing}, vol.~14, no.~6, pp. 1072--1087,
  2020.

\bibitem{yoo2021time}
J.~Yoo, K.~H. Jin, H.~Gupta, J.~Yerly, M.~Stuber, and M.~Unser,
  ``Time-dependent deep image prior for dynamic mri,'' \emph{IEEE Transactions
  on Medical Imaging}, vol.~40, no.~12, pp. 3337--3348, 2021.

\bibitem{kleineisel2022real}
J.~Kleineisel, J.~F. Heidenreich, P.~Eirich, N.~Petri, H.~K{\"o}stler,
  B.~Petritsch, T.~A. Bley, and T.~Wech, ``Real-time cardiac mri using an
  undersampled spiral k-space trajectory and a reconstruction based on a
  variational network,'' \emph{Magnetic Resonance in Medicine}, vol.~88, no.~5,
  pp. 2167--2178, 2022.

\bibitem{zhang2024camp}
L.~Zhang, X.~Li, and W.~Chen, ``Camp-net: Consistency-aware multi-prior network
  for accelerated mri reconstruction,'' \emph{IEEE Journal of Biomedical and
  Health Informatics}, pp. 1--14, 2024.

\bibitem{phair2024motion}
A.~Phair, A.~Fotaki, L.~Felsner, T.~J. Fletcher, H.~Qi, R.~M. Botnar, and
  C.~Prieto, ``A motion-corrected deep-learning reconstruction framework for
  accelerating whole-heart magnetic resonance imaging in patients with
  congenital heart disease,'' \emph{Journal of Cardiovascular Magnetic
  Resonance}, vol.~26, no.~1, p. 101039, 2024.

\bibitem{catalan2025unsupervised}
T.~Catal{\'a}n, M.~Courdurier, A.~Osses, A.~Fotaki, R.~Botnar,
  F.~Sahli-Costabal, and C.~Prieto, ``Unsupervised reconstruction of
  accelerated cardiac cine mri using neural fields,'' \emph{Computers in
  Biology and Medicine}, vol. 185, p. 109467, 2025.

\bibitem{liu2019mantis}
F.~Liu, L.~Feng, and R.~Kijowski, ``Mantis: model-augmented neural network with
  incoherent k-space sampling for efficient mr parameter mapping,''
  \emph{Magnetic resonance in medicine}, vol.~82, no.~1, pp. 174--188, 2019.

\bibitem{liu2020high}
F.~Liu, R.~Kijowski, L.~Feng, and G.~El~Fakhri, ``High-performance rapid mr
  parameter mapping using model-based deep adversarial learning,''
  \emph{Magnetic resonance imaging}, vol.~74, pp. 152--160, 2020.

\bibitem{liu2021magnetic}
F.~Liu, R.~Kijowski, G.~El~Fakhri, and L.~Feng, ``Magnetic resonance parameter
  mapping using model-guided self-supervised deep learning,'' \emph{Magnetic
  resonance in medicine}, vol.~85, no.~6, pp. 3211--3226, 2021.

\bibitem{jun2021deep}
Y.~Jun, H.~Shin, T.~Eo, T.~Kim, and D.~Hwang, ``Deep model-based magnetic
  resonance parameter mapping network (dopamine) for fast t1 mapping using
  variable flip angle method,'' \emph{Medical Image Analysis}, vol.~70, p.
  102017, 2021.

\bibitem{bian2024improving}
W.~Bian, A.~Jang, and F.~Liu, ``Improving quantitative mri using
  self-supervised deep learning with model reinforcement: Demonstration for
  rapid t1 mapping,'' \emph{Magnetic Resonance in Medicine}, vol.~92, no.~1,
  pp. 98--111, 2024.

\bibitem{jun2024zero}
Y.~Jun, Y.~Arefeen, J.~Cho, S.~Fujita, X.~Wang, P.~E. Grant, B.~Gagoski,
  C.~Jaimes, M.~S. Gee, and B.~Bilgic, ``Zero-deepsub: Zero-shot deep subspace
  reconstruction for rapid multiparametric quantitative mri using 3d-qalas,''
  \emph{Magnetic Resonance in Medicine}, vol.~91, no.~6, pp. 2459--2482, 2024.

\bibitem{Ho2020ddpm}
J.~Ho, A.~Jain, and P.~Abbeel, ``Denoising diffusion probabilistic models,''
  \emph{Advances in neural information processing systems}, vol.~33, pp.
  6840--6851, 2020.

\bibitem{nichol2021improved}
A.~Q. Nichol and P.~Dhariwal, ``Improved denoising diffusion probabilistic
  models,'' \emph{International Conference on Machine Learning}, pp.
  8162--8171, 2021.

\bibitem{dhariwal2021diffusion}
P.~Dhariwal and A.~Nichol, ``Diffusion models beat gans on image synthesis,''
  \emph{Advances in neural information processing systems}, vol.~34, pp.
  8780--8794, 2021.

\bibitem{chung2022score}
H.~Chung and J.~C. Ye, ``Score-based diffusion models for accelerated mri,''
  \emph{Medical image analysis}, vol.~80, p. 102479, 2022.

\bibitem{chung2022come}
H.~Chung, B.~Sim, and J.~C. Ye, ``Come-closer-diffuse-faster: Accelerating
  conditional diffusion models for inverse problems through stochastic
  contraction,'' \emph{Proceedings of the IEEE/CVF Conference on Computer
  Vision and Pattern Recognition}, pp. 12\,413--12\,422, 2022.

\bibitem{gungor2023adaptive}
A.~G{\"u}ng{\"o}r, S.~U. Dar, {\c{S}}.~{\"O}zt{\"u}rk, Y.~Korkmaz, H.~A. Bedel,
  G.~Elmas, M.~Ozbey, and T.~{\c{C}}ukur, ``Adaptive diffusion priors for
  accelerated mri reconstruction,'' \emph{Medical image analysis}, vol.~88, p.
  102872, 2023.

\bibitem{luo2023bayesian}
G.~Luo, M.~Blumenthal, M.~Heide, and M.~Uecker, ``Bayesian mri reconstruction
  with joint uncertainty estimation using diffusion models,'' \emph{Magnetic
  Resonance in Medicine}, vol.~90, no.~1, pp. 295--311, 2023.

\bibitem{bian2024diffusion}
W.~Bian, A.~Jang, L.~Zhang, X.~Yang, Z.~Stewart, and F.~Liu, ``Diffusion
  modeling with domain-conditioned prior guidance for accelerated mri and qmri
  reconstruction,'' \emph{IEEE Transactions on Medical Imaging}, 2024.

\bibitem{lin2025diffbir}
X.~Lin, J.~He, Z.~Chen, Z.~Lyu, B.~Dai, F.~Yu, Y.~Qiao, W.~Ouyang, and C.~Dong,
  ``Diffbir: Toward blind image restoration with generative diffusion prior,''
  \emph{European Conference on Computer Vision}, pp. 430--448, 2025.

\bibitem{huang2005k}
F.~Huang, J.~Akao, S.~Vijayakumar, G.~R. Duensing, and M.~Limkeman, ``k-t
  grappa: A k-space implementation for dynamic mri with high reduction
  factor,'' \emph{Magnetic Resonance in Medicine: An Official Journal of the
  International Society for Magnetic Resonance in Medicine}, vol.~54, no.~5,
  pp. 1172--1184, 2005.

\bibitem{zhang2023k}
L.~Zhang and W.~Chen, ``k-t clair: Self-consistency guided multi-prior learning
  for dynamic parallel mr image reconstruction,'' \emph{International Workshop
  on Statistical Atlases and Computational Models of the Heart}, pp. 314--325,
  2023.

\bibitem{feng2014golden}
L.~Feng, R.~Grimm, K.~T. Block, H.~Chandarana, S.~Kim, J.~Xu, L.~Axel, D.~K.
  Sodickson, and R.~Otazo, ``Golden-angle radial sparse parallel mri:
  combination of compressed sensing, parallel imaging, and golden-angle radial
  sampling for fast and flexible dynamic volumetric mri,'' \emph{Magnetic
  resonance in medicine}, vol.~72, no.~3, pp. 707--717, 2014.

\bibitem{feng2016xd}
L.~Feng, L.~Axel, H.~Chandarana, K.~T. Block, D.~K. Sodickson, and R.~Otazo,
  ``Xd-grasp: golden-angle radial mri with reconstruction of extra motion-state
  dimensions using compressed sensing,'' \emph{Magnetic resonance in medicine},
  vol.~75, no.~2, pp. 775--788, 2016.

\bibitem{winkelmann2006optimal}
S.~Winkelmann, T.~Schaeffter, T.~Koehler, H.~Eggers, and O.~Doessel, ``An
  optimal radial profile order based on the golden ratio for time-resolved
  mri,'' \emph{IEEE transactions on medical imaging}, vol.~26, no.~1, pp.
  68--76, 2006.

\bibitem{spincemaille2011z}
P.~Spincemaille, J.~Liu, T.~Nguyen, M.~R. Prince, and Y.~Wang, ``Z
  intensity-weighted position self-respiratory gating method for free-breathing
  3d cardiac cine imaging,'' \emph{Magnetic resonance imaging}, vol.~29, no.~6,
  pp. 861--868, 2011.

\bibitem{seiberlich2007non}
N.~Seiberlich, F.~A. Breuer, M.~Blaimer, K.~Barkauskas, P.~M. Jakob, and M.~A.
  Griswold, ``Non-cartesian data reconstruction using grappa operator gridding
  (grog),'' \emph{Magnetic Resonance in Medicine: An Official Journal of the
  International Society for Magnetic Resonance in Medicine}, vol.~58, no.~6,
  pp. 1257--1265, 2007.

\bibitem{seiberlich2008self}
N.~Seiberlich, F.~Breuer, M.~Blaimer, P.~Jakob, and M.~Griswold,
  ``Self-calibrating grappa operator gridding for radial and spiral
  trajectories,'' \emph{Magnetic Resonance in Medicine: An Official Journal of
  the International Society for Magnetic Resonance in Medicine}, vol.~59,
  no.~4, pp. 930--935, 2008.

\bibitem{benkert2018optimization}
T.~Benkert, Y.~Tian, C.~Huang, E.~V. DiBella, H.~Chandarana, and L.~Feng,
  ``Optimization and validation of accelerated golden-angle radial sparse mri
  reconstruction with self-calibrating grappa operator gridding,''
  \emph{Magnetic resonance in medicine}, vol.~80, no.~1, pp. 286--293, 2018.

\bibitem{wang2024cmrxrecon}
C.~Wang, J.~Lyu, S.~Wang, C.~Qin, K.~Guo, X.~Zhang, X.~Yu, Y.~Li, F.~Wang,
  J.~Jin \emph{et~al.}, ``Cmrxrecon: A publicly available k-space dataset and
  benchmark to advance deep learning for cardiac mri,'' \emph{Scientific Data},
  vol.~11, no.~1, p. 687, 2024.

\bibitem{zhang2013coil}
T.~Zhang, J.~M. Pauly, S.~S. Vasanawala, and M.~Lustig, ``Coil compression for
  accelerated imaging with cartesian sampling,'' \emph{Magnetic resonance in
  medicine}, vol.~69, no.~2, pp. 571--582, 2013.

\bibitem{glorot2010understanding}
X.~Glorot and Y.~Bengio, ``Understanding the difficulty of training deep
  feedforward neural networks,'' \emph{Proceedings of the thirteenth
  international conference on artificial intelligence and statistics}, pp.
  249--256, 2010.

\bibitem{loshchilov2017decoupled}
I.~Loshchilov, ``Decoupled weight decay regularization,'' \emph{arXiv preprint
  arXiv:1711.05101}, 2017.

\bibitem{krotkov1989active}
E.~P. Krotkov, ``Active computer vision by cooperative focus and stereo,''
  1989.

\bibitem{feng2020grasp}
L.~Feng, Q.~Wen, C.~Huang, A.~Tong, F.~Liu, and H.~Chandarana, ``Grasp-pro:
  improving grasp dce-mri through self-calibrating subspace-modeling and
  contrast phase automation,'' \emph{Magnetic resonance in medicine}, vol.~83,
  no.~1, pp. 94--108, 2020.

\bibitem{bian2025multi}
W.~Bian, A.~Jang, and F.~Liu, ``Multi-task magnetic resonance imaging
  reconstruction using meta-learning,'' \emph{Magnetic Resonance Imaging}, vol.
  116, p. 110278, 2025.

\bibitem{rombach2022high}
R.~Rombach, A.~Blattmann, D.~Lorenz, P.~Esser, and B.~Ommer, ``High-resolution
  image synthesis with latent diffusion models,'' \emph{Proceedings of the
  IEEE/CVF conference on computer vision and pattern recognition}, pp.
  10\,684--10\,695, 2022.

\bibitem{sandino2021deep}
C.~M. Sandino, F.~Ong, S.~S. Iyer, A.~Bush, and S.~Vasanawala, ``Deep subspace
  learning for efficient reconstruction of spatiotemporal imaging data,''
  \emph{NeurIPS 2021 Workshop on Deep Learning and Inverse Problems}, 2021.

\end{thebibliography}

\vfill\pagebreak
\section*{Supporting information}
The following supporting information is available as part of the online article:

\vskip\baselineskip\noindent
\textbf{Figure S1.}
{Qualitative comparison of different reconstruction methods in spatial and spatiotemporal dimensions, accompanied by corresponding error maps, for cardiac cine in long-axis views: two-chamber (2CH), three-chamber (3CH), and four-chamber (4CH). Spatiotemporal profiles along the yellow and red dotted lines are highlighted within yellow and red rectangles. Results are shown for an undersampling rate of 8$\times$. The proposed method demonstrates superior performance in recovering fine spatial details and preserving temporal dynamics.}

\noindent
\textbf{Figure S2.}
{Qualitative comparison of different reconstruction methods in spatial and spatiotemporal dimensions, accompanied by corresponding error maps, for cardiac cine in long-axis views: two-chamber (2CH), three-chamber (3CH), and four-chamber (4CH). Spatiotemporal profiles along the yellow and red dotted lines are highlighted within yellow and red rectangles. Results are presented for an undersampling rate of 10$\times$. The proposed method demonstrates superior performance in recovering fine spatial details and maintaining temporal dynamics.}

\noindent
\textbf{Figure S3.}
{Results of the ablation study for cine cardiac MR image reconstruction at 4$\times$ acceleration using dDiMo, demonstrating the impact of the weighting factors $\lambda_{xt}$ and $\lambda_{kt}$ on the loss function, with the noise estimation loss fixed at 1. Violin plots summarize reconstruction performance under the following conditions:
(A) Varying $\lambda_{xt}$ values while $\lambda_{kt}$ is fixed at 1.
(B) Varying $\lambda_{kt}$ values while $\lambda_{xt}$ is fixed at 1.
(C) Both $\lambda_{xt}$ and $\lambda_{kt}$ set to equal, varying values.
These experiments evaluate the relative contributions of $x$-$t$ and $k$-$t$ priors to the reconstruction of cine short-axis (SAX) and long-axis (LAX) cardiac MR images. For the experiments presented in the manuscript, $\lambda_{xt}$ and $\lambda_{kt}$ were both set to 1, achieving an optimal trade-off between noise estimation, $x$-$t$ prior, and $k$-$t$ prior learning, resulting in high reconstruction quality and efficiency.}

\noindent
\textbf{Figure S4.}
{Quality assessment results of dDiMo at 4$\times$ acceleration with varying total reverse diffusion steps during inference. Increasing the total diffusion steps improves the reconstructed images by enhancing sharpness and detail restoration. For the experiments presented in the main manuscript, 1000 diffusion steps were used, providing a balance between image quality and inference time efficiency.}

\noindent
\textbf{Figure S5.}
{Results of the ablation study validating the effectiveness of the conjugate gradient (CG) module in dDiMo during inference on continuously acquired golden-angle radial multicoil free-breathing lung data with 70 spokes. The study evaluates the impact of varying the weighting factor $\lambda_{td}$ in the objective function, which regulates the temporal finite difference constraint to enforce temporal sparsity. Tested $\lambda_{td}$ values range from 15 to 0.00015, including a condition where the CG module is excluded. Representative results are shown for Motion States 1, 2, and 6. The findings indicate that $\lambda_{td} = 0.015$ yields the best results across motion states, producing sharper images with finer details. Larger or smaller $\lambda_{td}$ values, as well as excluding the CG module, result in blurrier and noisier reconstructions. For the experiments presented in the main manuscript, $\lambda_{td} = 0.015$ was chosen to achieve an optimal trade-off between image quality and inference time efficiency.}

\noindent
\textbf{Figure S6.}
{3D U-Net architecture used for the noise estimation network $\epsilon_{\theta}$. The input $y_{t}$ has dimensions (H, W, T, 2C), where H and W represent the spatial height and width, T denotes the number of time frames, and C corresponds to the number of coils, with real and imaginary components stacked. The network employs 3D convolution kernels of size 3$\times$3$\times$3 and is designed as an encoder-decoder architecture with bottleneck blocks. Each encoder stage includes two residual blocks, with convolutional downsampling applied exclusively to the spatial dimensions (H and W), except at the final level. Corresponding decoder stages comprise three residual blocks and use 2$\times$ nearest-neighbor upsampling followed by convolutions to process inputs from the previous level. Skip connections between encoder and decoder stages ensure efficient feature propagation and preserve spatial and temporal details across the network. The model integrates ``Self-Attention" modules, indicated by red arrows, to enhance global contextual understanding. Additionally, each residual block incorporates inputs from the preceding layer and time-step embeddings, providing information about the current diffusion time step.}

\noindent
\textbf{Figure S7.}
{3D U-Net architecture used for the $x$-$t$ network $\Psi_{\theta}$. The input $\hat{y}_{0|t}$ has dimensions (H, W, T, 2C), where H and W represent the spatial height and width, T denotes the number of time frames, and C corresponds to the number of coils, with real and imaginary components stacked. The network employs 3D convolution kernels of size 3$\times$3$\times$3 and is structured as an encoder-decoder architecture with bottleneck blocks. Each encoder stage comprises two residual blocks, with convolutional downsampling applied exclusively to the spatial dimensions (H and W), except at the final level. The decoder stages contain three residual blocks and use 2$\times$ nearest-neighbor upsampling followed by convolutions to process inputs from the previous level. Skip connections between encoder and decoder stages ensure efficient feature propagation and preserve both spatial and temporal details. The model integrates ``Self-Attention" modules, indicated by red arrows, to enhance global contextual understanding. Additionally, each residual block receives inputs from the preceding layer and time-step embeddings, encoding information about the current diffusion time step. This design enables the network to effectively capture and process dynamic $x$-$t$ features, ensuring high spatial and temporal fidelity in the reconstructed outputs.}

\renewcommand{\figurename}{SUPPORTING INFORMATION FIGURE} 
\renewcommand{\thefigure}{S\arabic{figure}} 
\setcounter{figure}{0} 

\begin{figure*}[h]
	\centering
	\includegraphics[width=\textwidth]{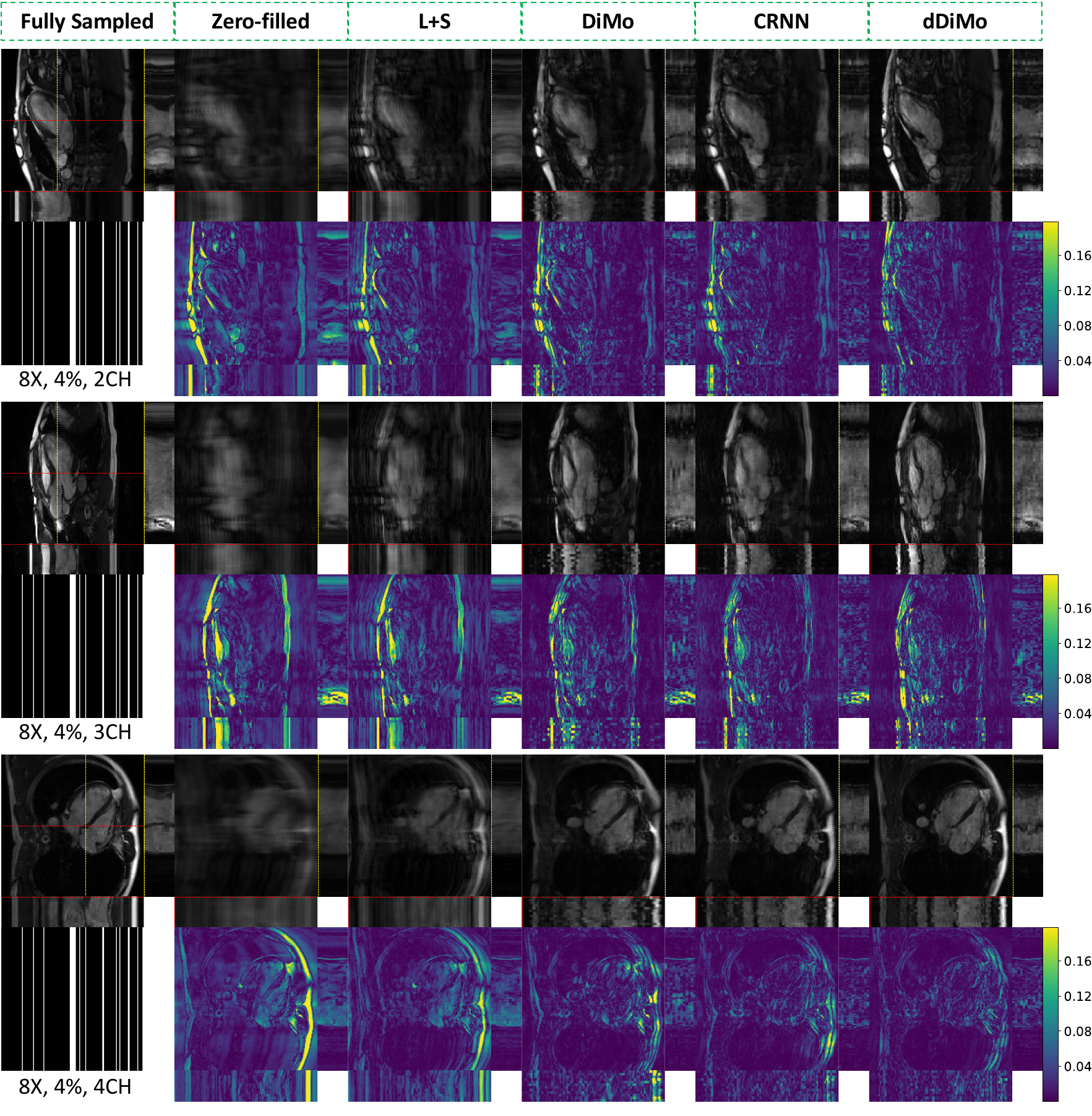}
	\caption{
        Qualitative comparison of different reconstruction methods in spatial and spatiotemporal dimensions, accompanied by corresponding error maps, for cardiac cine in long-axis views: two-chamber (2CH), three-chamber (3CH), and four-chamber (4CH). Spatiotemporal profiles along the yellow and red dotted lines are highlighted within yellow and red rectangles. Results are shown for an undersampling rate of 8$\times$. The proposed method demonstrates superior performance in recovering fine spatial details and preserving temporal dynamics.
    }
    \label{fig:cine-lax-8X}
\end{figure*}

\begin{figure*}[]
	\centering
	\includegraphics[width=\textwidth]{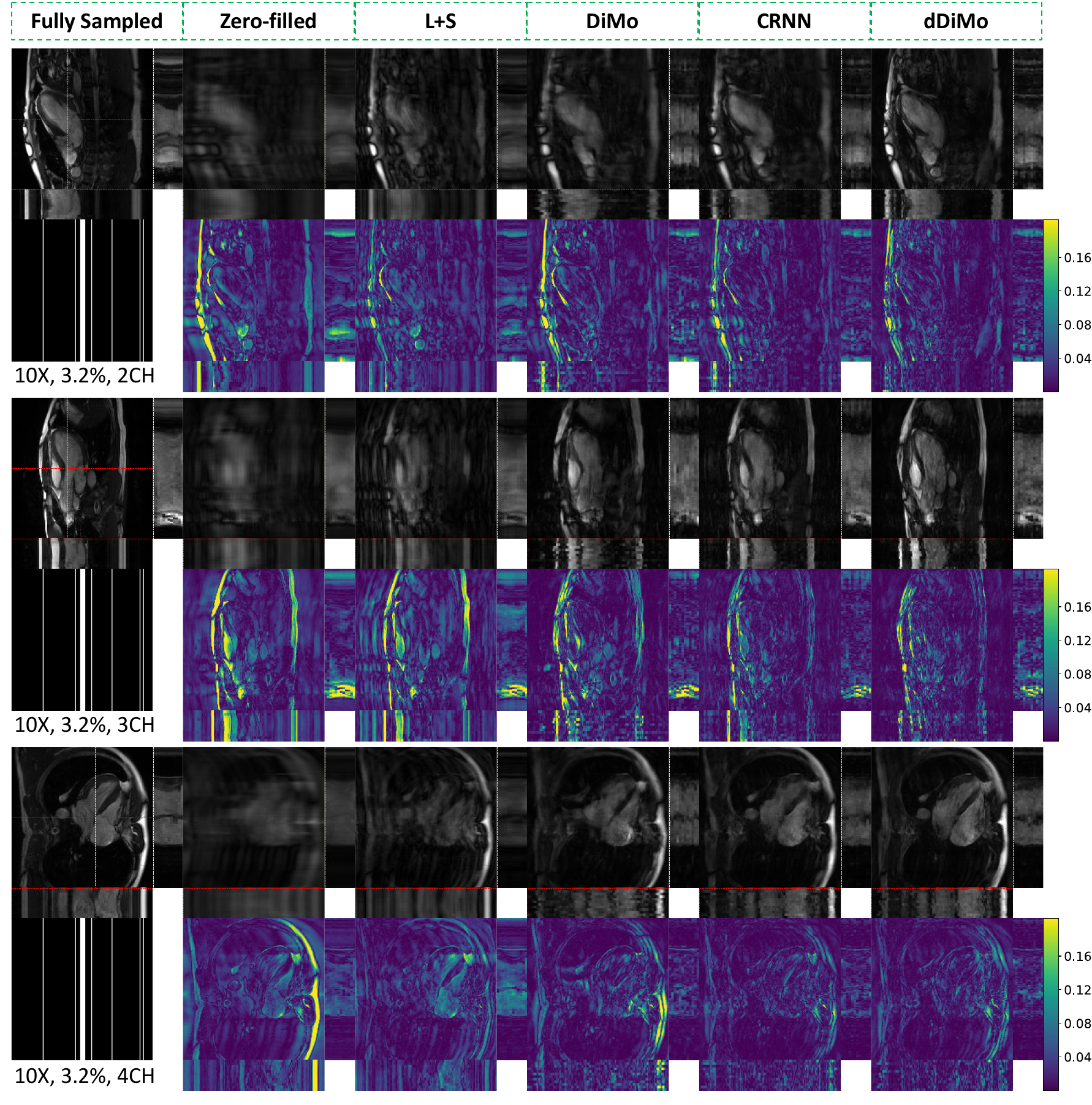}
	\caption{
        Qualitative comparison of different reconstruction methods in spatial and spatiotemporal dimensions, accompanied by corresponding error maps, for cardiac cine in long-axis views: two-chamber (2CH), three-chamber (3CH), and four-chamber (4CH). Spatiotemporal profiles along the yellow and red dotted lines are highlighted within yellow and red rectangles. Results are presented for an undersampling rate of 10$\times$. The proposed method demonstrates superior performance in recovering fine spatial details and maintaining temporal dynamics.
    }
    \label{fig:cine-lax-10X}
\end{figure*}

\begin{figure*}[]
    \centering
    \includegraphics[width=\linewidth]{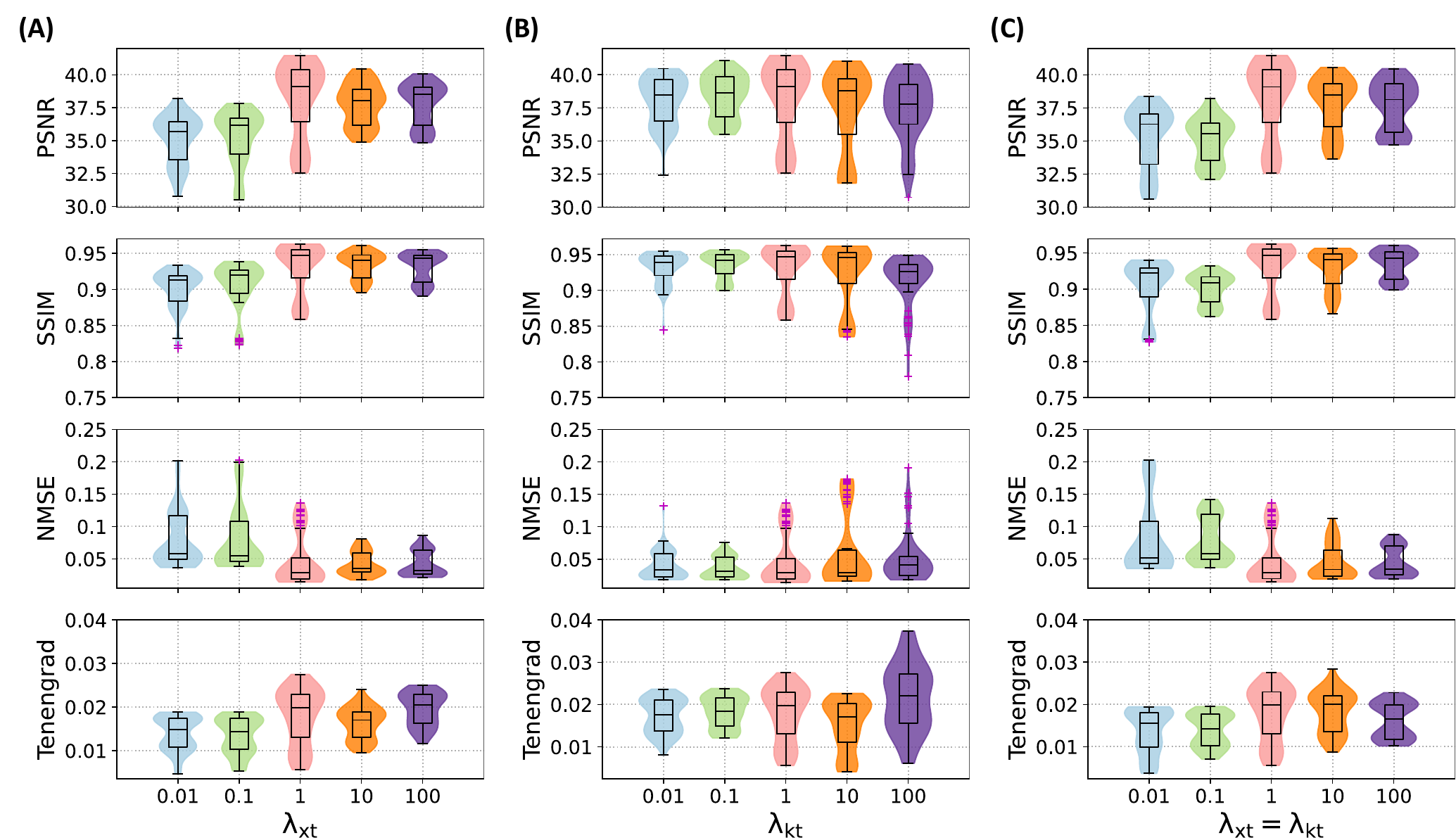}
    \caption{
        Results of the ablation study for cine cardiac MR image reconstruction at 4$\times$ acceleration using dDiMo, demonstrating the impact of the weighting factors $\lambda_{xt}$ and $\lambda_{kt}$ on the loss function, with the noise estimation loss fixed at 1. Violin plots summarize reconstruction performance under the following conditions:
        (A) Varying $\lambda_{xt}$ values while $\lambda_{kt}$ is fixed at 1.
        (B) Varying $\lambda_{kt}$ values while $\lambda_{xt}$ is fixed at 1.
        (C) Both $\lambda_{xt}$ and $\lambda_{kt}$ set to equal, varying values.
        These experiments evaluate the relative contributions of $x$-$t$ and $k$-$t$ priors to the reconstruction of cine short-axis (SAX) and long-axis (LAX) cardiac MR images. For the experiments presented in the manuscript, $\lambda_{xt}$ and $\lambda_{kt}$ were both set to 1, achieving an optimal trade-off between noise estimation, $x$-$t$ prior, and $k$-$t$ prior learning, resulting in high reconstruction quality and efficiency.
    }
\end{figure*}

\begin{figure*}[]
    \centering
    \includegraphics[width=\linewidth]{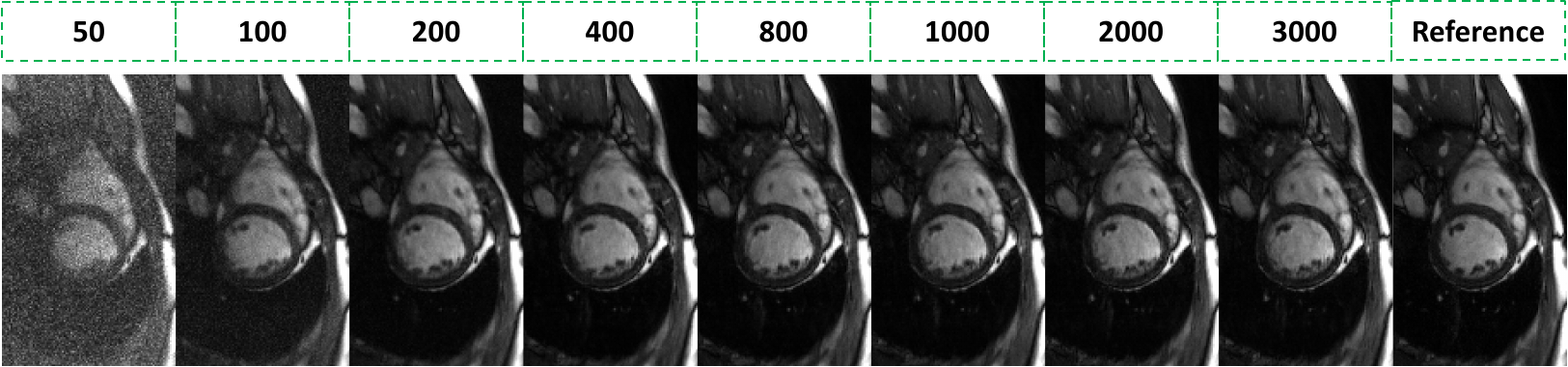}
    \caption{
    Quality assessment results of dDiMo at 4$\times$ acceleration with varying total reverse diffusion steps during inference. Increasing the total diffusion steps improves the reconstructed images by enhancing sharpness and detail restoration. For the experiments presented in the main manuscript, 1000 diffusion steps were used, providing a balance between image quality and inference time efficiency.
    }
\end{figure*}

\begin{figure*}[]
    \centering
    \includegraphics[width=\linewidth]{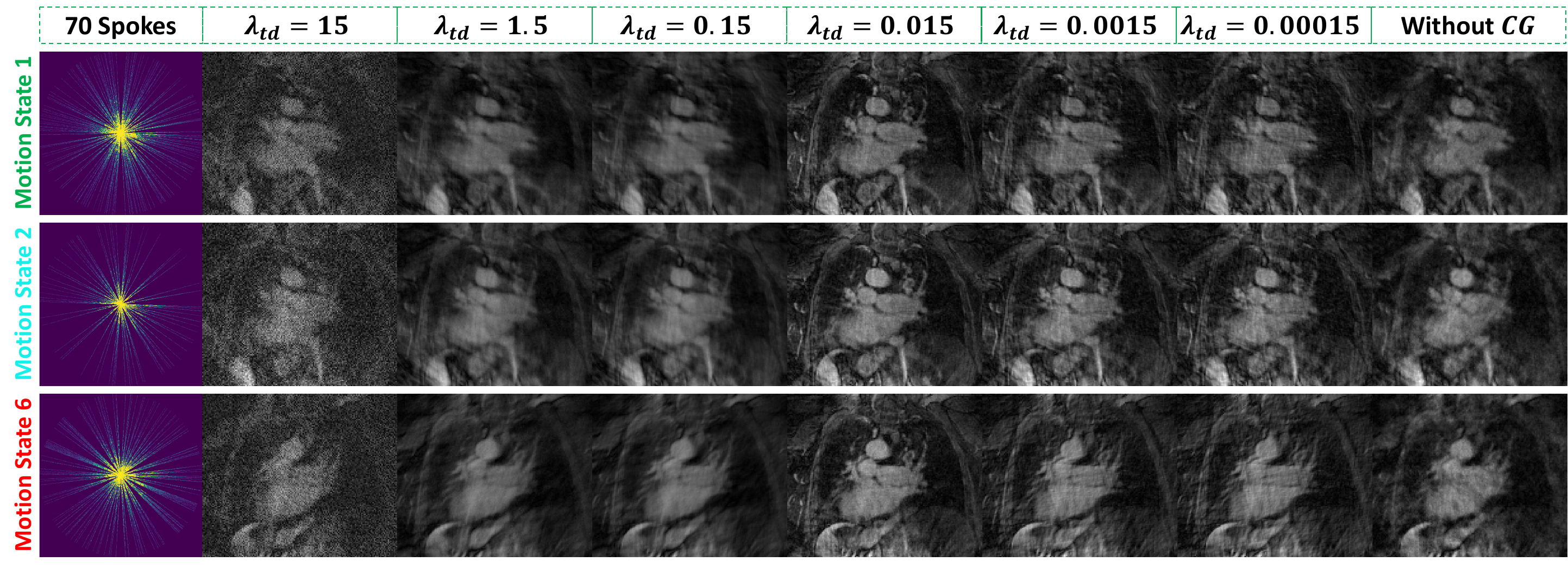}
    \caption{
        Results of the ablation study validating the effectiveness of the conjugate gradient (CG) module in dDiMo during inference on continuously acquired golden-angle radial multicoil free-breathing lung data with 70 spokes. The study evaluates the impact of varying the weighting factor $\lambda_{td}$ in the objective function, which regulates the temporal finite difference constraint to enforce temporal sparsity. Tested $\lambda_{td}$ values range from 15 to 0.00015, including a condition where the CG module is excluded. Representative results are shown for Motion States 1, 2, and 6. The findings indicate that $\lambda_{td} = 0.015$ yields the best results across motion states, producing sharper images with finer details. Larger or smaller $\lambda_{td}$ values, as well as excluding the CG module, result in blurrier and noisier reconstructions. For the experiments presented in the main manuscript, $\lambda_{td} = 0.015$ was chosen to achieve an optimal trade-off between image quality and inference time efficiency.
    }
\end{figure*}

\begin{figure*}[]
    \centering
    \includegraphics[width=\linewidth]{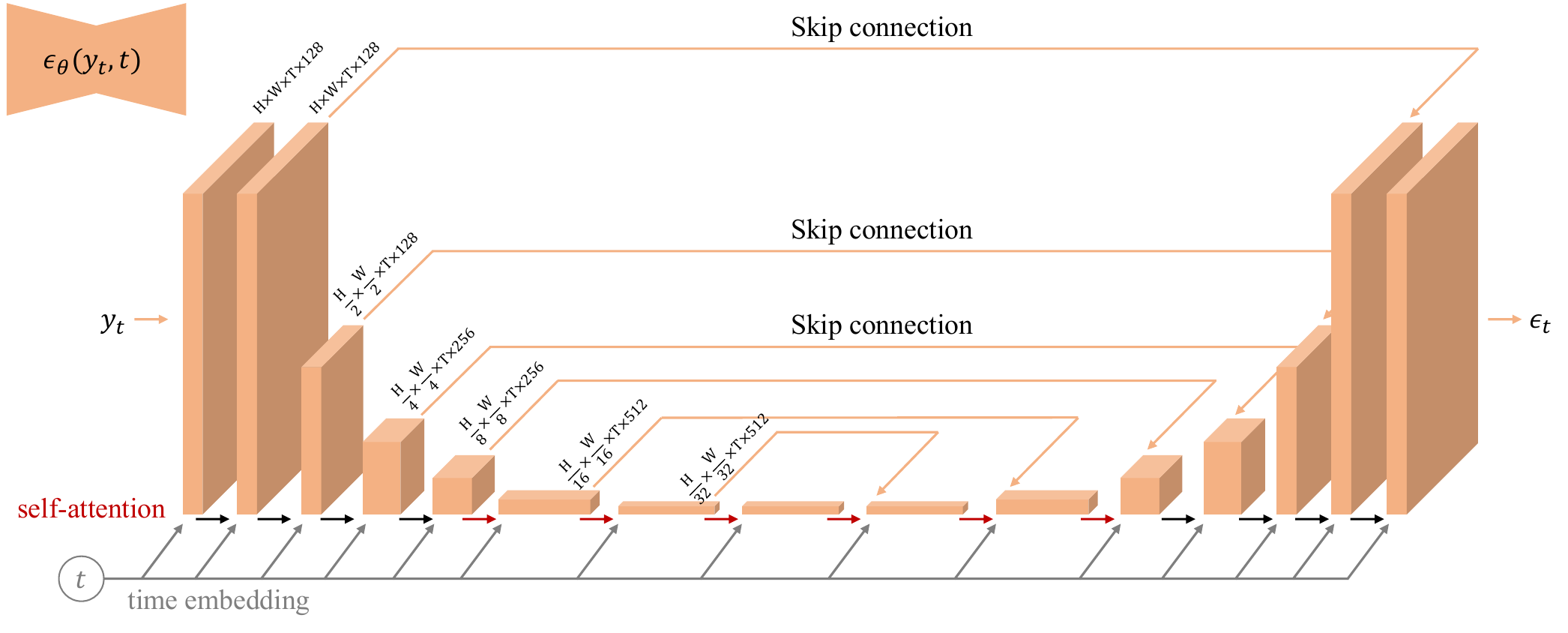}
    \caption{
        3D U-Net architecture used for the noise estimation network $\epsilon_{\theta}$. The input $y_{t}$ has dimensions (H, W, T, 2C), where H and W represent the spatial height and width, T denotes the number of time frames, and C corresponds to the number of coils, with real and imaginary components stacked. The network employs 3D convolution kernels of size 3$\times$3$\times$3 and is designed as an encoder-decoder architecture with bottleneck blocks. Each encoder stage includes two residual blocks, with convolutional downsampling applied exclusively to the spatial dimensions (H and W), except at the final level. Corresponding decoder stages comprise three residual blocks and use 2$\times$ nearest-neighbor upsampling followed by convolutions to process inputs from the previous level. Skip connections between encoder and decoder stages ensure efficient feature propagation and preserve spatial and temporal details across the network. The model integrates ``Self-Attention" modules, indicated by red arrows, to enhance global contextual understanding. Additionally, each residual block incorporates inputs from the preceding layer and time-step embeddings, providing information about the current diffusion time step.
    }
\end{figure*}

\begin{figure*}[]
    \centering
    \includegraphics[width=\linewidth]{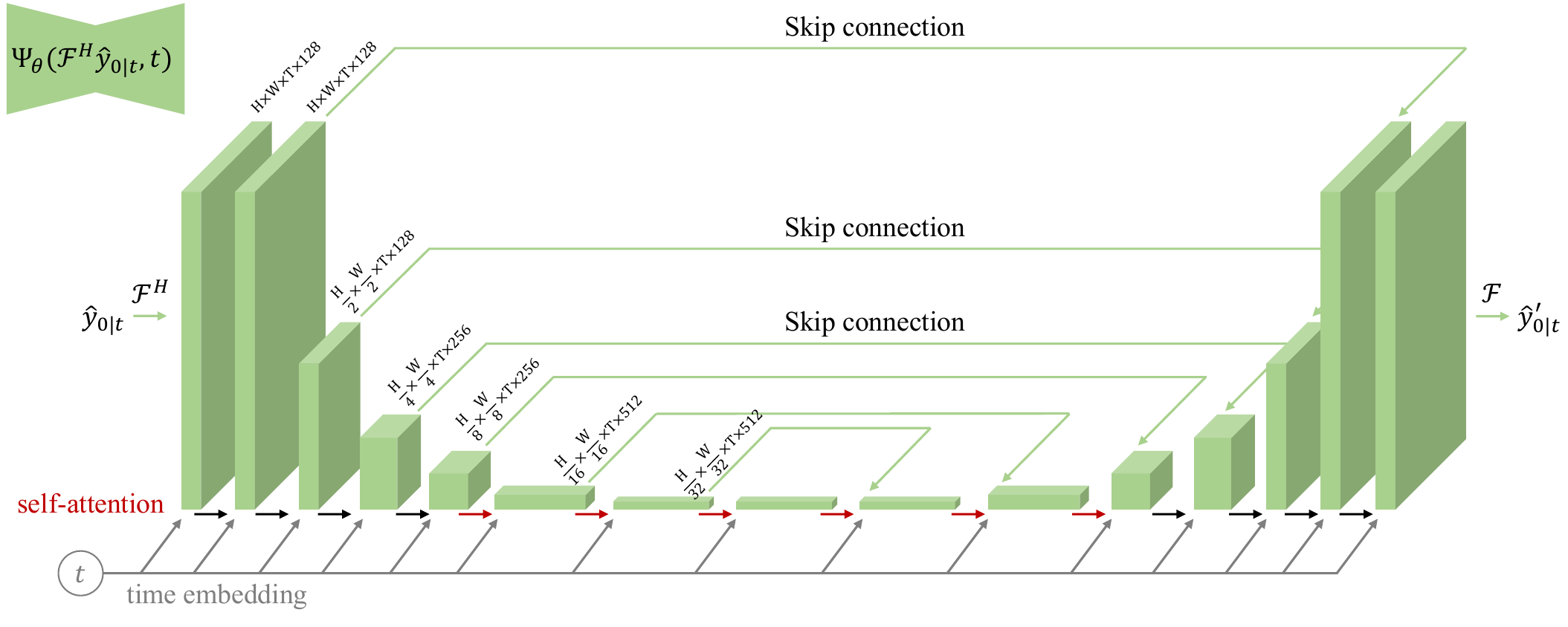}
    \caption{
        3D U-Net architecture used for the $x$-$t$ network $\Psi_{\theta}$. The input $\hat{y}_{0|t}$ has dimensions (H, W, T, 2C), where H and W represent the spatial height and width, T denotes the number of time frames, and C corresponds to the number of coils, with real and imaginary components stacked. The network employs 3D convolution kernels of size 3$\times$3$\times$3 and is structured as an encoder-decoder architecture with bottleneck blocks. Each encoder stage comprises two residual blocks, with convolutional downsampling applied exclusively to the spatial dimensions (H and W), except at the final level. The decoder stages contain three residual blocks and use 2$\times$ nearest-neighbor upsampling followed by convolutions to process inputs from the previous level. Skip connections between encoder and decoder stages ensure efficient feature propagation and preserve both spatial and temporal details. The model integrates ``Self-Attention" modules, indicated by red arrows, to enhance global contextual understanding. Additionally, each residual block receives inputs from the preceding layer and time-step embeddings, encoding information about the current diffusion time step. This design enables the network to effectively capture and process dynamic $x$-$t$ features, ensuring high spatial and temporal fidelity in the reconstructed outputs.
    }
\end{figure*}

\end{document}